\documentclass[superscriptaddress,amsmath,amssymb,aps,prl,floatfix, twocolumn, portrait]{revtex4-2}

\usepackage[english]{babel}
\usepackage{graphicx}
\usepackage{float}
\usepackage{dblfloatfix}
\usepackage{dcolumn}
\usepackage{bm}
\usepackage{physics}
\usepackage{siunitx}
\usepackage{dsfont}
\usepackage{booktabs}
\usepackage{comment}
\usepackage{xcolor}
\usepackage{todonotes}
\usepackage[mathlines]{lineno}
\usepackage{lscape} 
\usepackage{pdfpages}
\usepackage{pgffor}
\usepackage{etoolbox} 
\makeatletter
\AtBeginDocument{\let\LS@rot\@undefined}
\makeatother
\begin{document}

\title[Quantum State Interferography]{Quantum State Interferography}
\author{Surya Narayan Sahoo}
\affiliation{ 
Light and Matter Physics, Raman Research Institute, Bengaluru 560080, India
}
\author{Sanchari Chakraborti}
\affiliation{ 
Light and Matter Physics, Raman Research Institute, Bengaluru 560080, India
}
\author{Arun K. Pati}
\affiliation{
Quantum Information and Computation Group, Harish-Chandra Research Institute, HBNI, Allahabad 211019, India
}
\author{Urbasi Sinha}
 \email{usinha@rri.res.in}
\affiliation{ Light and Matter Physics, Raman Research Institute, Bengaluru 560080, India
}
\date{\today}
\begin{abstract}
Quantum State Tomography (QST) has been the traditional method for characterization of an unknown state. Recently, many direct measurement methods have been implemented to reconstruct the state in a resource efficient way. In this letter, we present an interferometric method, in which, any qubit state, whether mixed or pure, can be inferred from the visibility, phase shift and average intensity of an interference pattern using a single shot measurement- hence, we call it Quantum State Interferography. This provides us with a ``black box" approach to quantum state estimation, wherein, between the incidence of the photon and extraction of state information, we are not changing any conditions within the set-up, thus giving us a true single shot estimation of the quantum state. In contrast, standard QST requires at least two measurements for pure state qubit and at least three measurements for mixed state qubit reconstruction. We then go on to show that QSI is more resource efficient than QST for quantification of entanglement in pure bipartite qubits. We experimentally implement our method with high fidelity using the polarisation degree of freedom of light. An extension of the scheme to pure states involving $d-1$ interferograms for $d$-dimensional systems is also presented. Thus, the scaling gain is even more dramatic in the qudit scenario for our method where in contrast, standard QST, without any assumptions, scales roughly as $d^2$.

\end{abstract}

\maketitle

\section{\label{sec:Intro} Introduction}
The inherent probabilistic features of quantum measurement play a central
role in quantum mechanics. The probability distribution of outcomes of any
measurement on a quantum system can be predicted if its quantum state is
known. However, an unknown quantum
state of a single particle cannot be directly determined in
any experiment \cite{QSD_Barnett_2009}. Nevertheless, if we have an ensemble
of identically prepared particles, we can reconstruct the
quantum state by measuring the expectation values of different
observables.

 One of the widely used methods for state reconstruction is the Quantum State Tomography (QST) technique \cite{QSTtut_toninelli_2019, QSTfocus_banaszek_2013}, which often requires additional post-processing to ensure the physicality of the reconstructed density matrix \cite{QSTqubits_James_2001, MLE_Banaszek_1999}. For a $d$-dimensional system, typically one requires $d^{2} - 1$ measurements to reconstruct an arbitrary state. For a pure qudit state, measurement of $5d-7$ observables suffices to give us a unique state \cite{pureQST_Chen_2013, pauliQST_Ma_2016}. Over the last decade, several schemes towards improving the scaling of QST with the dimension of the Hilbert space have been suggested \cite{efficient_Cramer_2010, LRE_qi_2013, QSTnob_Diaz_2015} and recently the focus has been towards single-shot state estimation, i.e., obtaining the state in a single-set up without any required change in the experimental settings \cite{ss-seq_diLorenzo_2013, ss-pulse_lin_2016, ss-trans_zhu_2019, ss-geo_athira_2019}. 

In this letter, we present a novel method for reconstructing (pure or mixed) quantum state of a qubit along with its experimental implementation, and also extend the scheme to infer the state of $d$-dimensional qudits requiring only $d-1$ measurements, which serves as a promising and less cumbersome alternative to QST. We emphasise that in our proposed scheme, the number of measurements scales linearly with the dimensionality of the system whereas, in general, the required number of measurements for QST scales quadratically with respect to the system size. Thus, for higher dimensional systems, our method is more economical compared to QST. Our method can also be used for quantification and reconstruction of bipartite pure entangled states in an efficient manner.

 Earlier, other alternatives to standard QST using projective measurements have been explored, in which the strength of interaction may be strong as in \cite{alt-seq_diLorenzo_2013} or weak as in \cite{weakQST_wu_2013, alt-weak_Chen_2018, alt-opt_Shojaee_2018}. Since weak measurements \cite{wm_Aharonov_1988, wm_Duck_1989} can give us complex weak values of observables, they have paved the way for direct measurement of quantum state \cite{dm-wf_Lundeen_2011, dm-procedure_Lundeen_2012, dm-dm_thekkadath_2016, dm-pol_salvail_2013, dm27_Malik_2014,dm-compressive_mirhosseini_2014,dm1M_Zhimin_2016}. Our work in this letter focuses on the use of interferometric method as opposed to direct measurement techniques to obtain the quantum state.

 Recently, it has been shown by us \cite{non_hermitian_expt_Nirala_2019} and others \cite{ anomalousweakvalues_Abbott_2019}
that complex weak values can be obtained without performing weak measurement, which can lead to efficient direct measurement of quantum states \cite{wv-framework_Ogawa_2019}. Knowing the weak value of a Hermitian operator can give us the expectation value of related non-Hermitian operator \cite{non_hermitian_theory_Pati_2015}.
Expectation value of non-Hermitian column operators have been used for direct measurement of the state \cite{directNonHermitian_bolduc_2016}.

In this letter, we show that interferometric methods can be used to infer the quantum state of an ensemble of identically prepared qubits by a single-shot measurement \footnote{Here, by single-shot, we mean that the measurement settings do not change during the course of data acquisition}. We name our method Quantum State Interferography (QSI). QSI focuses on minimizing number of data acquisitions as all parameters describing the state are obtained at once by post-processing the interference pattern. This differs from direct state measurement which focuses on minimizing post-processing at the cost of changing the experimental set up. QSI has enormous practical benefits vis-a-vis quantum state estimation that can be useful in various applications like quantum information processing protocols \cite{photonicReview_Flamini_2018, photonicReview_slussarenko_2019}. We experimentally implement it in polarization degree of freedom of light, which yields a single-shot method for characterization of polarization state of light.

In the next section\ref{sec:Theory_Qubits}, we discuss the theory for how a two-path interferometer can be used to reconstruct not only pure states but also mixed states. We then experimentally demonstrate the method using \num{632.8} \si{\nano \metre} Helium-Neon laser light in a displaced Sagnac interferometer 
\cite{displacedSagnac_Micuda_2014}. Then\ref{sec:qudits}, we extend the protocol to qudits and show the advantage of using QSI over QST for pure states.

\section{\label{sec:Theory_Qubits} Quantum State Interferography for qubits} 
The general density matrix for a qubit can be written using the coordinates $\theta \in [0, \pi]$ and  $\phi \in (-\pi, \pi]$, that describe the direction of the vector in the Bloch sphere representation and $\mu \in [0,1]$, which is related to the purity of the state and the length of the vector.
\begin{align}
	\rho = 
	\begin{pmatrix}
		\cos^2 \left(\frac{\theta}{2} \right) & \frac{1}{2} \mu e^{-i \phi} \sin(\theta) \\
		\frac{1}{2} \mu e^{i \phi} \sin(\theta) & \sin^2 \left(\frac{\theta}{2} \right)
	\end{pmatrix}.
	\label{eqn_rho}
\end{align}
The expectation value of spin-ladder operators $\sigma_{\pm} = \frac{1}{2}(\sigma_{x}\pm i \sigma_{y})$ is given as
\begin{align}
    \ev{\sigma_{\pm}} = \Tr(\rho \sigma_{\pm}) = \frac{1}{2} \exp(\pm i \phi) \ \mu \sin(\theta) \ . 
\end{align}
The argument of the complex expectation value $\ev{\sigma_{\pm}}$ directly gives us the azimuthal coordinate, i.e., $\phi = \pm \arg (\ev{\sigma_{\pm}}) $.   
For a pure state, $\mu=1$ and hence, the $\theta$ can be obtained as $\sin^{-1}(2 \abs{\ev{\sigma_{\pm}}})$. However, the solution to $\theta$ is not unique in $[0, \pi]$ and $\pi-\theta$ is a solution as well. Thus, to uniquely determine $\theta$, we need to measure the expectation value of another column operator, which in this case is the projector $\Pi_{0} = \dyad{0}$, with  
{$\ev{\Pi_0} = \cos^2(\theta/2)$}.Once $\ev{\Pi_0}$ is known, $\theta$ is uniquely determined in $[0, \pi]$. Now, $\mu$ can be determined as $\mu = \frac{2 \abs{\ev{\sigma_{\pm}}}}{\sin(\theta)}$.

\begin{figure}
    \centering
    \includegraphics[width = 0.99\linewidth]{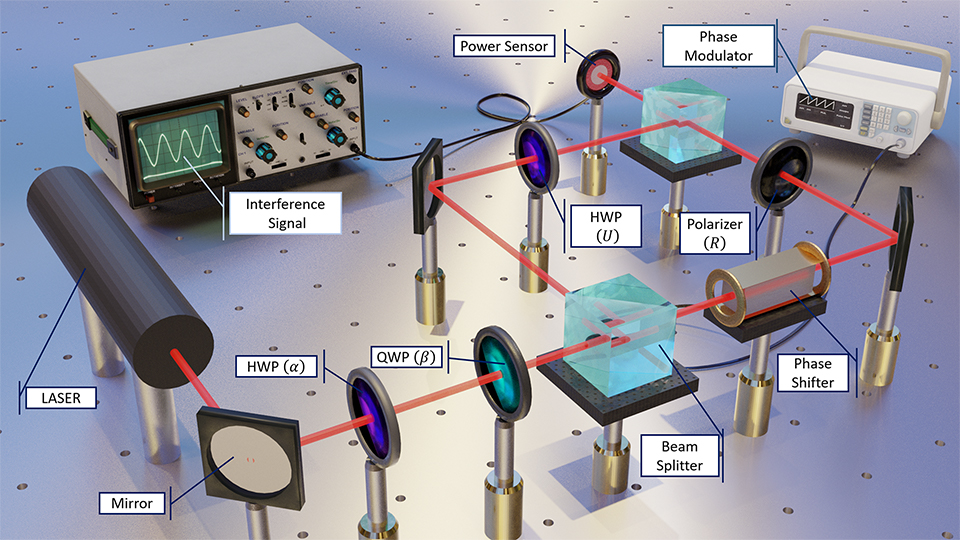}
    \caption{The polarization state is prepared by using half-wave plate (HWP) and quarter-wave plate (QWP) which can be at arbitrary orientations. The MZI is formed by the two beam splitters (BS). On one arm we place a HWP oriented at $\pi/4$ to realize the $\sigma_x$ operator. On the other arm, we use a polarizer with transmission axis oriented along horizontal, or alternatively, the transmitting port of the polarizing beam splitter (PBS) to effectively realize the operator $\Pi_{0}$. The phase shifter (PS) introduces a relative phase $\varphi$ between the two arms and we measure the intensity at the photodetector (PD) as a function of $\varphi$. Experimentally, the phase shifter can be avoided by making the interferometer non-collinear (See Fig. \ref{fig:Expt_Sagnac_SetUp} and \cite{non_hermitian_expt_Nirala_2019}) to obtain the interferogram in a single-shot.}
    \label{fig:qubit_mzi}
\end{figure}

 Next, we show that all three quantities $\theta$, $\phi$ and $\mu$ specifying the polarization state of light can be determined from a single interference pattern obtained in the MZI as shown in Fig. \ref{fig:qubit_mzi}.
The operator $A=\sigma_{-}$ is polar decomposed into the  $R=\Pi_{0}$ and $U = \sigma_x$. The optical components corresponding to $R$ and $U$ are placed in each arm of the MZI. 

We will discuss the scheme by taking an example of a MZI. However, it can
be realized with any two-path interferometer including
double-slit interferometer which can be factory designed
and can serve as a robust miniature device for state estimation.
If a pure state $\ket{\psi}$ is incident onto the first beam splitter (BS) of a MZI , the intensity at the photodetector \cite{non_hermitian_expt_Nirala_2019} is given by

\begin{align}
I(\varphi) = \frac{1}{4} ( 1 + \ev{\Pi_{0}} 
   + 2\abs{\ev{\sigma_{-}}} \cos(\arg{(\ev{\sigma_{-}})}+\varphi)) \ .
\end{align}
By knowing $I(\varphi)$, which can be experimentally obtained from a single interference pattern, we can determine $\ev{\Pi_0}$ and $\ev{\sigma_{-}}$.

If the incident state is a mixed state given by $\rho$, we obtain the intensity at the detector \footnote{See Supplementary for derivation} as follows:
\begin{align}
	I(\varphi) = \frac{1}{8} (
	3+ \cos(\theta) + 2 \mu  \sin (\theta ) \cos (\varphi -\phi))  .
\end{align}
The phase shift of the interference pattern is obtained at the value of $\varphi$ that maximizes $I(\varphi)$.
Since $0 \leq \theta \leq \pi \Rightarrow \sin(\theta)> 0$, the phase shift is obtained as $\Phi = \phi$. 

The phase averaged intensity and the visibility are given by 
 \begin{align}
\Bar{I}=\frac{1}{8} (3+\cos(\theta))
, \label{eqn:avgIntensity}  \qquad 
V = \frac{2 \mu \sin(\theta)}{3 + \cos(\theta)}
 \end{align}
where $\theta \in [0, \pi]$ can be uniquely determined  from $\Bar{I}$, which is experimentally always normalized with the incident intensity. Once $\theta$ is known, $\mu$ can be obtained from visibility and $\rho$ can be reconstructed.

Interestingly, QSI can also be used to quantify entanglement of pure bipartite states. If a bipartite state is pure, then entanglement can be quantified by the entanglement entropy - the von Neumman entropy of the reduced density matrix i.e., $E = - \Tr(\rho_A \log{(\rho_A)})$, where $\rho_A = \Tr_B(\rho_{AB})$ and $\rho_{AB} = \ket{\Psi}_{AB} \bra{\Psi}_{AB}$ \cite{entanglementEntropy_Bennet_1996,entanglementEntropy_Horodeki_2009, entanglementEntropy_Sorkin_1983}.
Since, with a single experimental set up the reduced density matrix $\rho_A$, which in general is a mixed state, can be determined using QSI, it can be used to quantify entanglement of pure states of bipartite qubits. The state $\ket{\Psi}_{AB}$ can be completely reconstructed with one additional interferogram as we have shown in accompanying supplementary material \footnote{The state reconstruction of bipartite qubit pure states is detailed in the \textit{Supplementary Material Sec.} $XIV$ }.

\section
{\label{sec:Expt} Experiment}
To reconstruct the state of various input polarizations we need to measure the phase shift of the interference pattern. If one uses a MZI, it needs to be phase-stabilized against vibrations that change the path difference. Thus, to avoid this, we prefer interferometers that are not prone to vibrations such as the Sagnac interferometer \footnote{ See supplementary for a comparison between interferometers for QSI, which includes Ref. [51] }. 
\begin{figure}[H]
	\centering
	\includegraphics[width=0.99\linewidth]{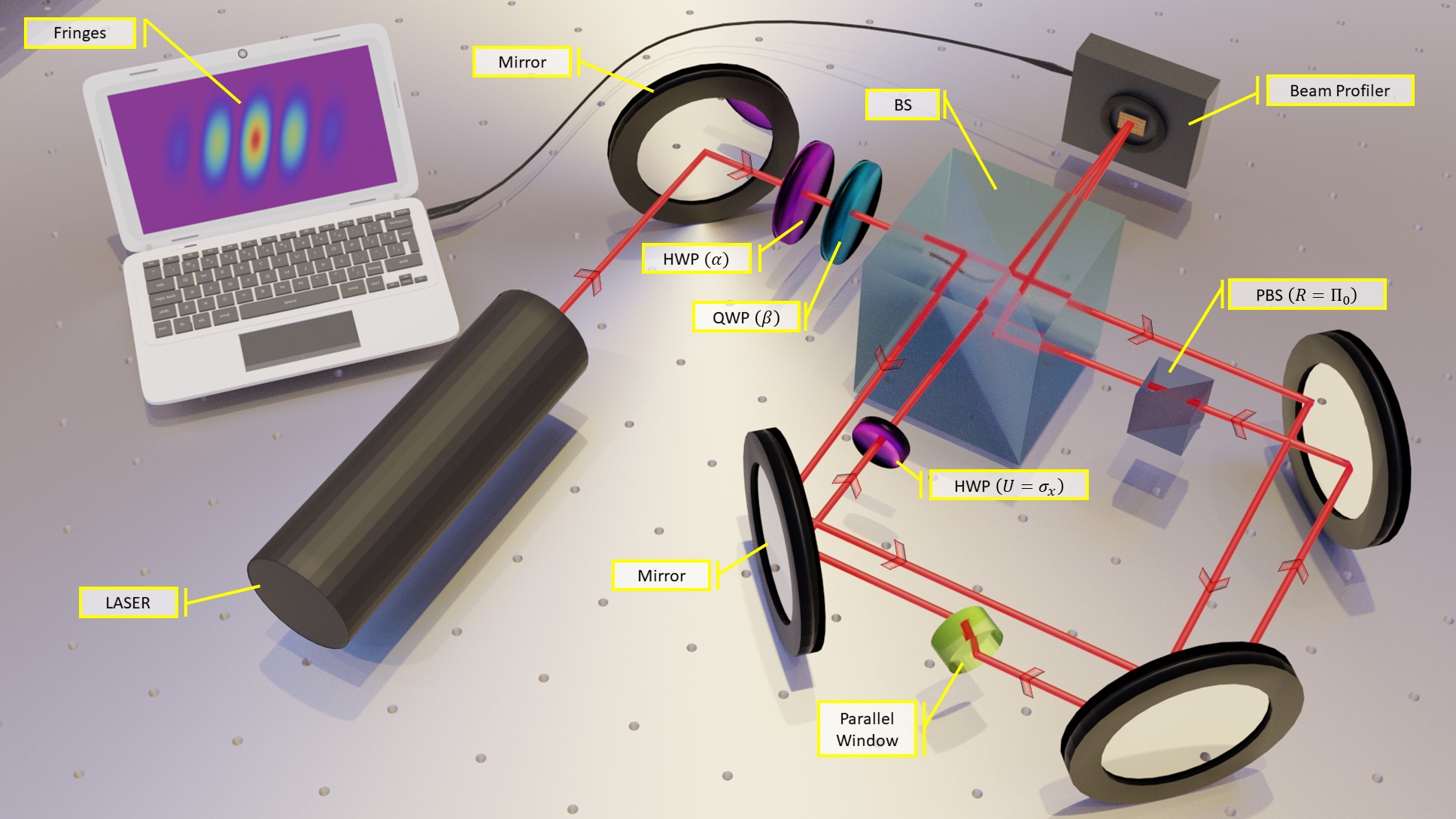}
	\caption{Non-Collinear displaced Sagnac Interferometer for Polarization State Interferography:  We use the Sagnac interferometer in non-collinear configuration \cite{non_hermitian_expt_Nirala_2019}, i.e., we tilt the beam splitter (Thorlabs BS013) to obtain double-slit like interference pattern. We use the displaced Sagnac configuration \cite{displacedSagnac_Micuda_2014} instead of the common-path configuration in order to place  the polarizing beam splitter (PBS, Thorlabs PBS122) in one arm (as $R=\Pi_0$ operation) and the HWP (Thorlabs WPH05M-633) in the other arm as the $U=\sigma_x$ operation. The  glass plate, (parallel window WG40530-B) placed in one of the paths is tilted to achieve the displacement of that beam to ensure maximum overlap of the two non-collinear beams at the beam profiler.}
	\label{fig:Expt_Sagnac_SetUp}
\end{figure}

\subsection{Methods}
The input state is prepared by placing a HWP (Thorlabs WPH05M-633) at an angle $\alpha$ followed by a QWP (Thorlabs WPQ05M-633) at angle $\beta$ in the path of a vertically polarized beam from a Helium Neon Laser (632.8 nm) before it enters the interferometer. 
For a fixed angle $\alpha$ of the HWP, we rotate the QWP and obtain 5 images for a given $\beta$. For each image, we take 100 horizontal slices about the vertical centroid and fit each slice with the model which is a Gaussian weighted cosine function:
\begin{align}
B_f + A_f e^{- c_f (x_f-m_f)^2} (1+v_f \cos(k_f x_f + \phi_f)) .
\label{fitFunctionNew}
\end{align}
Here, $B_f$ is the background noise, $A_f$ is the amplitude of the Gaussian envelope centred at $m_f$ with standard deviation  $\sqrt{1/2 c_f}$. The fringe width is given by $2\pi /k_f$. The visibility of the fringe and phase shift are determined from $v_f$ and $\phi_f$ respectively.\\
\\

\subsection{Results}

\paragraph{Phase Shift, Average Intensity and Visibility from the Interferogram}

\begin{figure}[H]
\centering
\includegraphics[width=0.99\linewidth]{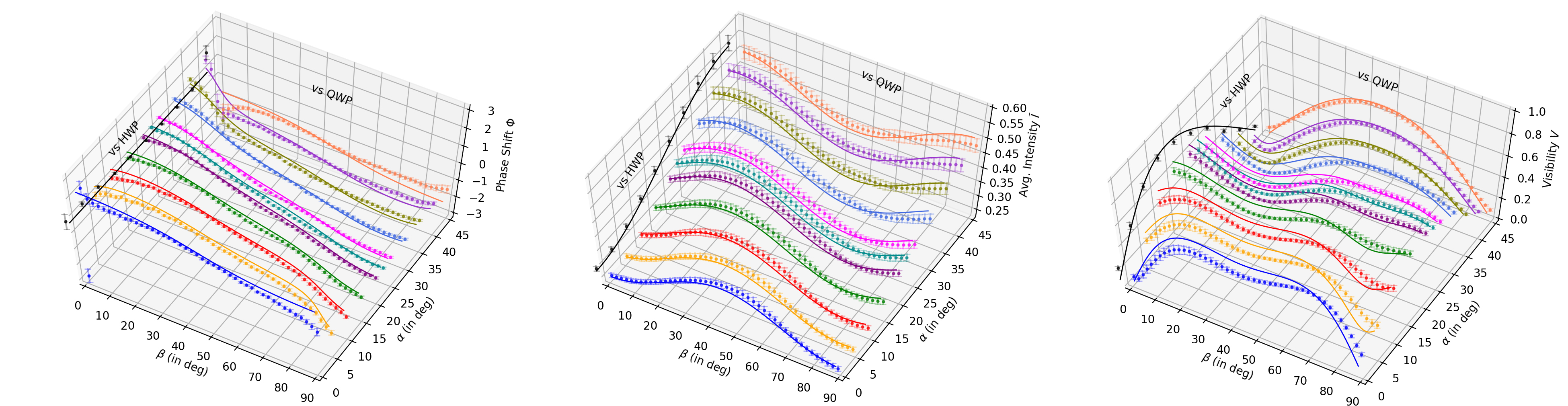}
\caption{Phase shift, Avg. Intensity and Visibility as a function of $\alpha$ and $\beta$. The solid lines in the plots represent the theoretical prediction while the dots and bars represent experimentally obtained mean and statistical error respectively. The black curve (in the $\beta=0$ plane) is for the experiment where only HWP was rotated in absence of the QWP.}
\label{fig:results3d}
\end{figure}

From the interference pattern obtained in the non-collinear displaced Sagnac interferometer, we determine the phase shift, the visibility and the average intensity for different polarization states prepared by different combination of HWP and QWP angles ($\alpha, \beta$) as shown in Fig. \ref{fig:results3d}.
 The error bars in the plots are obtained from statistics over the 100 slices for the 5 images.
 In absence of QWP, the experimentally obtained value of phase is expected to be a constant w.r.t $\alpha$. This is considered as the zero reference for all the measurements.
 The mean and standard deviations associated with phase were obtained from the experimental datasets using circular statistics \cite{circstats_fisher_1993}. The phase shift obtained from the interferogram has more error when $\theta$ is closer to $0$ or $\pi$, since the Bloch vector is closer to the poles where $\phi$ is undefined, which is manifested in noticeable deviations from the theory in the experimental plot for HWP angles $0^\circ$ and $45^\circ$. 

All the experimentally obtained averaged intensity are normalized (with norm = \num{0.5}) with respect to the corresponding maximum of the average intensity obtained as a function of HWP in the absence of QWP. The average intensity does not depend on the interference and hence is not prone to errors that affect the visibility and the phase shift. 
The experimentally obtained visibility is systematically lower than the theoretical prediction because of various experimental imperfections like polarization dependence of splitting ratio of the the beam splitter (about 3\%), angular deviation due to the rotation of the wave plates (10 arcsec)  that changes the spatial overlap and the intensity averaging over the pixel area.

\onecolumngrid

\twocolumngrid

\begin{figure}[H]
	\centering
 	\includegraphics[width=0.95\linewidth]{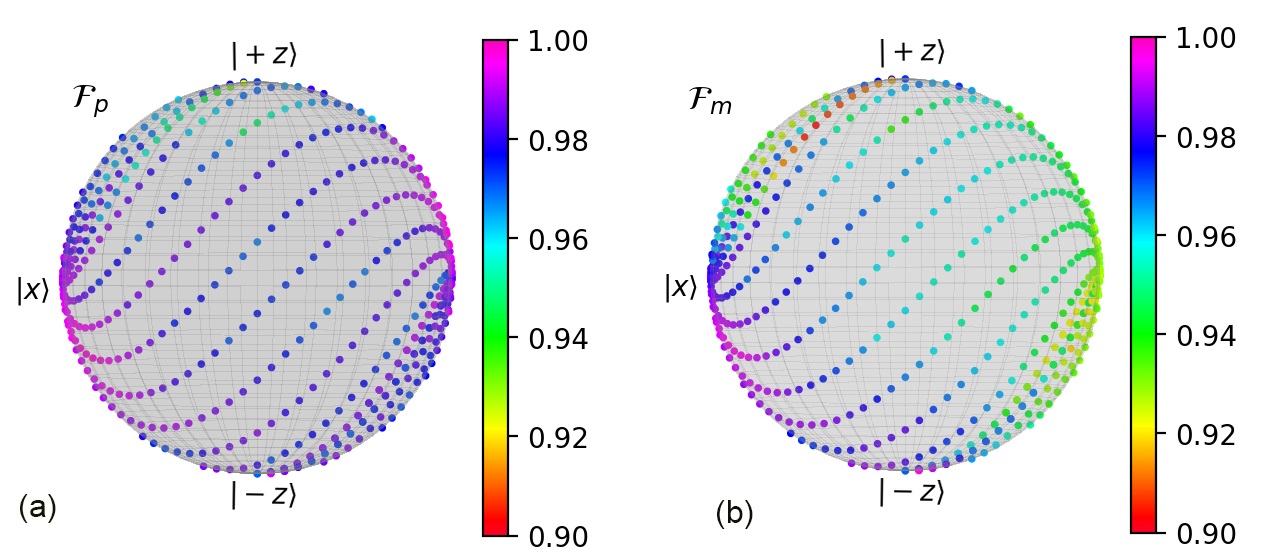}
	\caption{[Left] Fidelity with assumption that the various prepared states in $\theta$ and $\phi$ over the Bloch Sphere are pure. [Right] Fidelity of reconstructed density matrices of various prepared states in $\theta$ and $\phi$ over the Bloch Sphere.}
	\label{fig:Fidelity}
\end{figure}

\paragraph{Purity and Fidelity --}
 Assuming that the polarization state of the incident beam is pure, we compute the fidelity of the state reconstructed from $\theta$ and $\phi$ determined by the experimentally obtained average intensity $\Bar{I}$ and phase shift $\Phi$ respectively. The errors obtained in $\Phi$ and $\Bar{I}$ are propagated to the calculation of fidelity for a single state. The mean fidelity calculated from experimentally obtained mean phase shift ($\Phi$) and mean average intensity ($\Bar{I}$) are plotted on the Bloch sphere at the $\theta$ and $\phi$ of the prepared state in Fig. \ref{fig:Fidelity} (Left) with the values indicated by the colorbar. The average fidelity over all the prepared states is greater than 98 \% .

Although the incident state was almost pure ($>$ 99\% vertically polarized), our method can be used in experiments involving mixed states as well. To illustrate, we reconstruct the density matrix, as given in Eqn. (\ref{eqn_rho}) using the $\mu$ value determined from the experimentally obtained visibility, with the restriction that it makes the reconstructed density matrix physical, i.e., $\Tr(\rho^2) \leq 1$. 
This is ensured by construction of $\rho$ in Eqn. (\ref{eqn_rho}) with the restriction that we substitute $\mu$ with $\min{(\mu, 1)}$ as discussed in detail in accompanying supplementary material \footnote{The details of how $\mu$ affects the fidelity and other methods to obtain $\mu$ from experiment that makes $\rho$ physical is discussed in \textit{Supplementary Material Sec VIII}}.
%
%
Since the experimentally obtained visibility is systematically lower than the theory, the reconstructed state has a lower purity and consequently the fidelity  (in Fig. \ref{fig:Fidelity} [Right]) is lower than the case with pure state assumption.

\section{\label{sec:qudits} Quantum State Interferography\\ for Qudit pure states}
The pure state of a $d$-dimensional qudit can be represented in the polar spherical \cite{polar_Blumenson_1960} form as follows:
\begin{align}
    \ket{\psi}^{(d)} =
    \begin{pmatrix}
    \cos \left(\theta_1/2 \right) \\
    \sin \left(\theta_1 /2 \right) \exp(i \phi_1) \cos \left(\theta_2 / 2 \right) \\
    \vdots \\
    \prod_{j=1}^{k-1} \sin \left(\theta_j / 2 \right) \exp(i \phi_j) \cos \left(\theta_k / 2 \right) \\
    \vdots \\
    \prod_{j=1}^{d-1} \sin \left(\theta_j /2 \right) \exp(i \phi_j) 
    \end{pmatrix}.
\end{align}
The component of $\ket{\psi}^{(d)}$ in the $k$-th  2 dimensional subspace is given by
\begin{align}
\begin{aligned}
\ket{\psi}_{k}^{(2;d)} = 
\left(\prod_{j=1}^{k-1} \sin \left(\frac{\theta_j}{2}\right) e^{i \phi_j} \right)
\begin{pmatrix}
 \cos \left(\frac{\theta_{k}}{2} \right) \\
 \sin \left(\frac{\theta_{k}}{2} \right) e^{i \phi_{k}} \cos\left(\frac{\theta_{k+1}}{2}\right)
\end{pmatrix}
\ .
\end{aligned}
\end{align}

 We use $d-1$ interferometers, one on each of the two dimensional $\{k, k+1\}$ subspaces of the $d$-dimensional state $\ket{\psi}^{(d)}$. The expectation values 
of $\sigma_{-}$ and $\Pi_0$ operators for the $2$-dimensional subspace can be obtained directly from phase averaged intensity and phase shift of the interference pattern. Although, here we shall be formulating QSI for qudits using $d-1$ interferometers for ease of conceptualization, in principle and in many physical systems in practice, the state can be inferred from $d-1$ interferograms obtained with a setup involving only two interferometers. This is achieved by using the same interferometer for all the two dimensional subspaces ( please see Supplementary Material \footnote{Supplementary Material Sec XIII}).

The matrix element of the spin ladder operator \footnote{We shall use the notation $\mathcal{O}^{(k)}$ to denote the operator $\mathcal{O}$ meant for qubits realized in the $k$-th 2-dimensional subspace. The operators for $d$-dimensional qudits are represented as $\mathcal{O}^{[k]}$.} in the two dimensional subspace is
\begin{align}
\ev{\sigma_{\pm}^{(k)}}{\psi}_{k}^{(2;d)} = \xi(k)
\frac{1}{2} e^{\pm i \phi_k} \sin(\theta_{k}) \cos\left(\frac{\theta_{k+1}}{2}\right) 
\label{eqn: qudit_ladder_ev}
\end{align} 
where, $ \xi(k) = \prod_{j=1}^{k-1} \sin^2 \left(\frac{\theta_j}{2}\right)$.

We directly obtain the relative phase $\phi_k$ in the two-dimensional subspace from the argument of the matrix element of the spin ladder operator in that subspace. To determine $\theta_{k}$, however, we need to know $\xi(k)$ and $\theta_{k+1}$ as well. Nevertheless, as in the case for qubits, we need to measure the matrix element of $\Pi_{0}^{(k)}$ in the two-dimensional subspace, i.e.,
\begin{align}
\ev{\Pi_{0}^{(k)}}{\psi}_{k}^{(2;d)} = \xi(k) \cos^2\left(\frac{\theta_k}{2}\right).
 \label{eqn: qudit_projector_ev}
\end{align}

We can determine $\theta_{1}$ and subsequently $\theta_{2}$ as follows:
\begin{align}
    \frac{\theta_{1}}{2}=\cos^{-1}\left(\sqrt{\ev{\Pi_{0}^{(1)}}}\right), ~ \frac{\theta_{2}}{2}= \cos^{-1}\left(\frac{2 \abs{\ev{\sigma_{\pm}^{(1)}}}}{\sin(\theta_{1})}\right)
\end{align} 
Thus, once $\theta_k$ is determined, $\theta_{k+1}$ can be obtained sequentially.

 We can employ the same scheme to obtain $\ev{\sigma_{\pm}^{(k)}}$ by placing the polar decomposed elements $\Pi_{0}^{(k)}$ in one and $\sigma_x^{(k)}$ in the other arm of a MZI constructed for the 2-dimensional subspace $\{k, k+1\}$. We have to design $d-1$ such MZI setups for the state estimation of a $d$-dimensional qudit.

\begin{figure}[H]
	\centering
	\includegraphics[,  width=0.99\linewidth]{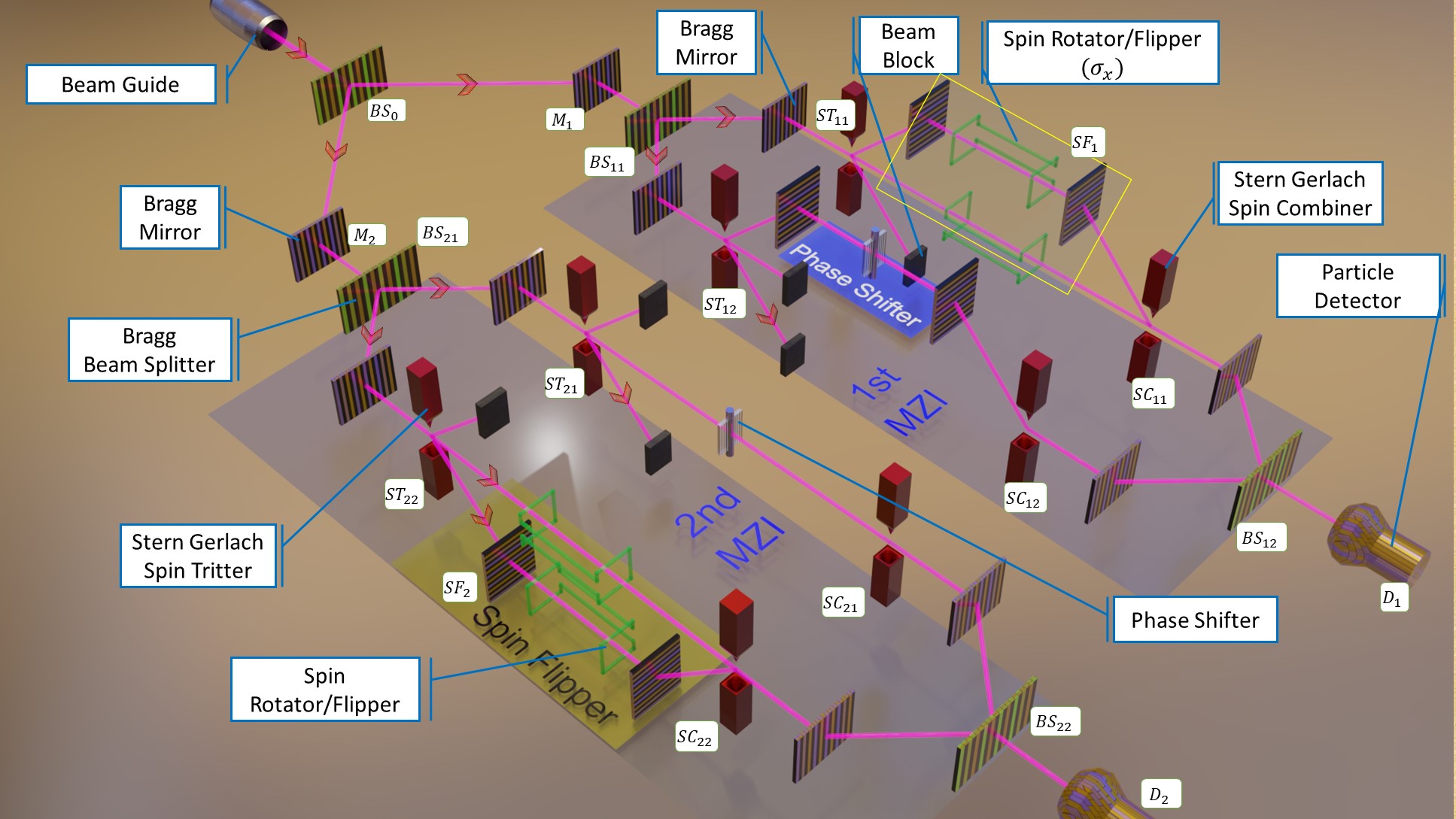}
	\caption{Schematic to measure the state for a qutrit, which can be generalized to any qudit: The beam is divided into d-1 spatial modes. Each mode is made incident on an interferomter with $\Pi_0^{(k)}$ and $\sigma_x^{(k)}$ operations corresponding to the $2$-dimensional subspace (For the detailed description see Supplementary Material Sec XI).} 
	\label{fig:qutritSchematic}
\end{figure}

Next, we present a generic scheme to construct all the necessary operators in each subspace from the Pauli operators in the $d$-dimensional Hilbert space. We illustrate the same using the example of qutrits in Fig \ref{fig:qutritSchematic}.

This scheme can be generalized with $d$-dimensions simply by blocking all other components after the spin splitter (ST) except the desired pair. Please see Supplementary Material Sec. XII for detailed expressions on how to infer the state from $d-1$ interferograms, which is also shown to be obtained with two interferometers in Sec. XIII.

\section{Conclusions and Outlook}
In summary, we have proposed quantum state interferography as a method to reconstruct a qubit state, whether pure or mixed, in a single experimental set up, and experimentally demonstrated our scheme with high average fidelity. 
This forms an efficient scheme compared to quantum state tomography as well as direct measurement techniques to infer the state of an ensemble of identically prepared qubits.

All the parameters needed to determine the state are obtained from the interference pattern produced using a single shot measurement. Since the interference pattern obtained using the coherent laser light source and a stream of single photons would be identical \cite{quantumClassicalEquiv_Skagerstam_2018, OET_Sudarshan_1963}, the method described here is applicable for determining state of identically prepared ensemble of single photons as well. 
QSI provides us with a ``black box" approach to quantum state estimation, wherein, between the incidence of the photon and extraction of state information, we are not changing any conditions within the set-up, which itself can be miniaturized . This provides us a true single shot estimation of the quantum state which has a rich potential for future technological development.

We have also shown here how QSI can be extended to estimate pure states of $d$-dimensional qudits with $d-1$ measurements, which can be obtained either by using $d-1$ interferometers as shown in Fig. \ref{fig:qutritSchematic} or by using only two interferometers as shown in accompanying supplementary material. This is achieved by representing a $d$-dimensional qudit using $2 (d-1)$ parameters and extracting 2 parameters from each interferogram. While for qubits, we require one measurement as opposed to three in standard QST, the improvement is even more tremendous for qudits where standard quantum state tomography, without any assumptions, scales roughly as $d^2$ and for pure states, the scaling has been brought down to $5d - 7$ \cite{pureQST_Chen_2013, pauliQST_Ma_2016} 
so far. This may help in efficient characterization of higher dimensional systems \cite{Spatial_Ghosh_2018} aimed towards quantum information processing, quantum computation and quantum communication.The QSI can also be used for single-shot entanglement quantification of pure bipartite states, which can be useful towards foundation of quantum mechanics.

\begin{acknowledgements}
U.S acknowledges the research grant from Department of Science and Technology under the QuEST network programme for partial support.
\end{acknowledgements}

\bibliography{main}

\begin{thebibliography}{51}%
\makeatletter
\providecommand \@ifxundefined [1]{%
 \@ifx{#1\undefined}
}%
\providecommand \@ifnum [1]{%
 \ifnum #1\expandafter \@firstoftwo
 \else \expandafter \@secondoftwo
 \fi
}%
\providecommand \@ifx [1]{%
 \ifx #1\expandafter \@firstoftwo
 \else \expandafter \@secondoftwo
 \fi
}%
\providecommand \natexlab [1]{#1}%
\providecommand \enquote  [1]{``#1''}%
\providecommand \bibnamefont  [1]{#1}%
\providecommand \bibfnamefont [1]{#1}%
\providecommand \citenamefont [1]{#1}%
\providecommand \href@noop [0]{\@secondoftwo}%
\providecommand \href [0]{\begingroup \@sanitize@url \@href}%
\providecommand \@href[1]{\@@startlink{#1}\@@href}%
\providecommand \@@href[1]{\endgroup#1\@@endlink}%
\providecommand \@sanitize@url [0]{\catcode `\\12\catcode `\$12\catcode
  `\&12\catcode `\#12\catcode `\^12\catcode `\_12\catcode `\%12\relax}%
\providecommand \@@startlink[1]{}%
\providecommand \@@endlink[0]{}%
\providecommand \url  [0]{\begingroup\@sanitize@url \@url }%
\providecommand \@url [1]{\endgroup\@href {#1}{\urlprefix }}%
\providecommand \urlprefix  [0]{URL }%
\providecommand \Eprint [0]{\href }%
\providecommand \doibase [0]{https://doi.org/}%
\providecommand \selectlanguage [0]{\@gobble}%
\providecommand \bibinfo  [0]{\@secondoftwo}%
\providecommand \bibfield  [0]{\@secondoftwo}%
\providecommand \translation [1]{[#1]}%
\providecommand \BibitemOpen [0]{}%
\providecommand \bibitemStop [0]{}%
\providecommand \bibitemNoStop [0]{.\EOS\space}%
\providecommand \EOS [0]{\spacefactor3000\relax}%
\providecommand \BibitemShut  [1]{\csname bibitem#1\endcsname}%
\let\auto@bib@innerbib\@empty
\bibitem [{\citenamefont {Barnett}\ and\ \citenamefont
  {Croke}(2009)}]{QSD_Barnett_2009}%
  \BibitemOpen
  \bibfield  {author} {\bibinfo {author} {\bibfnamefont {S.~M.}\ \bibnamefont
  {Barnett}}\ and\ \bibinfo {author} {\bibfnamefont {S.}~\bibnamefont
  {Croke}},\ }\bibfield  {title} {\bibinfo {title} {Quantum state
  discrimination},\ }\href {https://doi.org/10.1364/AOP.1.000238} {\bibfield
  {journal} {\bibinfo  {journal} {Advances in Optics and Photonics}\ }\textbf
  {\bibinfo {volume} {1}},\ \bibinfo {pages} {238} (\bibinfo {year}
  {2009})}\BibitemShut {NoStop}%
\bibitem [{\citenamefont {Toninelli}\ \emph {et~al.}(2019)\citenamefont
  {Toninelli}, \citenamefont {Ndagano}, \citenamefont {Vallés}, \citenamefont
  {Sephton}, \citenamefont {Nape}, \citenamefont {Ambrosio}, \citenamefont
  {Capasso}, \citenamefont {Padgett},\ and\ \citenamefont
  {Forbes}}]{QSTtut_toninelli_2019}%
  \BibitemOpen
  \bibfield  {author} {\bibinfo {author} {\bibfnamefont {E.}~\bibnamefont
  {Toninelli}}, \bibinfo {author} {\bibfnamefont {B.}~\bibnamefont {Ndagano}},
  \bibinfo {author} {\bibfnamefont {A.}~\bibnamefont {Vallés}}, \bibinfo
  {author} {\bibfnamefont {B.}~\bibnamefont {Sephton}}, \bibinfo {author}
  {\bibfnamefont {I.}~\bibnamefont {Nape}}, \bibinfo {author} {\bibfnamefont
  {A.}~\bibnamefont {Ambrosio}}, \bibinfo {author} {\bibfnamefont
  {F.}~\bibnamefont {Capasso}}, \bibinfo {author} {\bibfnamefont {M.~J.}\
  \bibnamefont {Padgett}},\ and\ \bibinfo {author} {\bibfnamefont
  {A.}~\bibnamefont {Forbes}},\ }\bibfield  {title} {\bibinfo {title} {Concepts
  in quantum state tomography and classical implementation with intense light:
  a tutorial},\ }\href {https://doi.org/10.1364/AOP.11.000067} {\bibfield
  {journal} {\bibinfo  {journal} {Advances in Optics and Photonics}\ }\textbf
  {\bibinfo {volume} {11}},\ \bibinfo {pages} {67} (\bibinfo {year}
  {2019})}\BibitemShut {NoStop}%
\bibitem [{\citenamefont {Banaszek}\ \emph {et~al.}(2013)\citenamefont
  {Banaszek}, \citenamefont {Cramer},\ and\ \citenamefont
  {Gross}}]{QSTfocus_banaszek_2013}%
  \BibitemOpen
  \bibfield  {author} {\bibinfo {author} {\bibfnamefont {K.}~\bibnamefont
  {Banaszek}}, \bibinfo {author} {\bibfnamefont {M.}~\bibnamefont {Cramer}},\
  and\ \bibinfo {author} {\bibfnamefont {D.}~\bibnamefont {Gross}},\ }\bibfield
   {title} {\bibinfo {title} {Focus on quantum tomography},\ }\href
  {https://doi.org/10.1088/1367-2630/15/12/125020} {\bibfield  {journal}
  {\bibinfo  {journal} {New Journal of Physics}\ }\textbf {\bibinfo {volume}
  {15}},\ \bibinfo {pages} {125020} (\bibinfo {year} {2013})}\BibitemShut
  {NoStop}%
\bibitem [{\citenamefont {James}\ \emph {et~al.}(2001)\citenamefont {James},
  \citenamefont {Kwiat}, \citenamefont {Munro},\ and\ \citenamefont
  {White}}]{QSTqubits_James_2001}%
  \BibitemOpen
  \bibfield  {author} {\bibinfo {author} {\bibfnamefont {D.~F.~V.}\
  \bibnamefont {James}}, \bibinfo {author} {\bibfnamefont {P.~G.}\ \bibnamefont
  {Kwiat}}, \bibinfo {author} {\bibfnamefont {W.~J.}\ \bibnamefont {Munro}},\
  and\ \bibinfo {author} {\bibfnamefont {A.~G.}\ \bibnamefont {White}},\
  }\bibfield  {title} {\bibinfo {title} {Measurement of qubits},\ }\href
  {https://doi.org/10.1103/PhysRevA.64.052312} {\bibfield  {journal} {\bibinfo
  {journal} {Physical Review A}\ }\textbf {\bibinfo {volume} {64}},\ \bibinfo
  {pages} {052312} (\bibinfo {year} {2001})}\BibitemShut {NoStop}%
\bibitem [{\citenamefont {Banaszek}\ \emph {et~al.}(1999)\citenamefont
  {Banaszek}, \citenamefont {D’Ariano}, \citenamefont {Paris},\ and\
  \citenamefont {Sacchi}}]{MLE_Banaszek_1999}%
  \BibitemOpen
  \bibfield  {author} {\bibinfo {author} {\bibfnamefont {K.}~\bibnamefont
  {Banaszek}}, \bibinfo {author} {\bibfnamefont {G.~M.}\ \bibnamefont
  {D’Ariano}}, \bibinfo {author} {\bibfnamefont {M.~G.~A.}\ \bibnamefont
  {Paris}},\ and\ \bibinfo {author} {\bibfnamefont {M.~F.}\ \bibnamefont
  {Sacchi}},\ }\bibfield  {title} {\bibinfo {title} {Maximum-likelihood
  estimation of the density matrix},\ }\href
  {https://doi.org/10.1103/PhysRevA.61.010304} {\bibfield  {journal} {\bibinfo
  {journal} {Physical Review A}\ }\textbf {\bibinfo {volume} {61}},\ \bibinfo
  {pages} {010304} (\bibinfo {year} {1999})}\BibitemShut {NoStop}%
\bibitem [{\citenamefont {Chen}\ \emph {et~al.}(2013)\citenamefont {Chen},
  \citenamefont {Dawkins}, \citenamefont {Ji}, \citenamefont {Johnston},
  \citenamefont {Kribs}, \citenamefont {Shultz},\ and\ \citenamefont
  {Zeng}}]{pureQST_Chen_2013}%
  \BibitemOpen
  \bibfield  {author} {\bibinfo {author} {\bibfnamefont {J.}~\bibnamefont
  {Chen}}, \bibinfo {author} {\bibfnamefont {H.}~\bibnamefont {Dawkins}},
  \bibinfo {author} {\bibfnamefont {Z.}~\bibnamefont {Ji}}, \bibinfo {author}
  {\bibfnamefont {N.}~\bibnamefont {Johnston}}, \bibinfo {author}
  {\bibfnamefont {D.}~\bibnamefont {Kribs}}, \bibinfo {author} {\bibfnamefont
  {F.}~\bibnamefont {Shultz}},\ and\ \bibinfo {author} {\bibfnamefont
  {B.}~\bibnamefont {Zeng}},\ }\bibfield  {title} {\bibinfo {title} {Uniqueness
  of quantum states compatible with given measurement results},\ }\href
  {https://doi.org/10.1103/PhysRevA.88.012109} {\bibfield  {journal} {\bibinfo
  {journal} {Phys. Rev. A}\ }\textbf {\bibinfo {volume} {88}},\ \bibinfo
  {pages} {012109} (\bibinfo {year} {2013})}\BibitemShut {NoStop}%
\bibitem [{\citenamefont {Ma}\ \emph {et~al.}(2016)\citenamefont {Ma},
  \citenamefont {Jackson}, \citenamefont {Zhou}, \citenamefont {Chen},
  \citenamefont {Lu}, \citenamefont {Mazurek}, \citenamefont {Fisher},
  \citenamefont {Peng}, \citenamefont {Kribs}, \citenamefont {Resch},
  \citenamefont {Ji}, \citenamefont {Zeng},\ and\ \citenamefont
  {Laflamme}}]{pauliQST_Ma_2016}%
  \BibitemOpen
  \bibfield  {author} {\bibinfo {author} {\bibfnamefont {X.}~\bibnamefont
  {Ma}}, \bibinfo {author} {\bibfnamefont {T.}~\bibnamefont {Jackson}},
  \bibinfo {author} {\bibfnamefont {H.}~\bibnamefont {Zhou}}, \bibinfo {author}
  {\bibfnamefont {J.}~\bibnamefont {Chen}}, \bibinfo {author} {\bibfnamefont
  {D.}~\bibnamefont {Lu}}, \bibinfo {author} {\bibfnamefont {M.~D.}\
  \bibnamefont {Mazurek}}, \bibinfo {author} {\bibfnamefont {K.~A.~G.}\
  \bibnamefont {Fisher}}, \bibinfo {author} {\bibfnamefont {X.}~\bibnamefont
  {Peng}}, \bibinfo {author} {\bibfnamefont {D.}~\bibnamefont {Kribs}},
  \bibinfo {author} {\bibfnamefont {K.~J.}\ \bibnamefont {Resch}}, \bibinfo
  {author} {\bibfnamefont {Z.}~\bibnamefont {Ji}}, \bibinfo {author}
  {\bibfnamefont {B.}~\bibnamefont {Zeng}},\ and\ \bibinfo {author}
  {\bibfnamefont {R.}~\bibnamefont {Laflamme}},\ }\bibfield  {title} {\bibinfo
  {title} {Pure-state tomography with the expectation value of pauli
  operators},\ }\href {https://doi.org/10.1103/PhysRevA.93.032140} {\bibfield
  {journal} {\bibinfo  {journal} {Phys. Rev. A}\ }\textbf {\bibinfo {volume}
  {93}},\ \bibinfo {pages} {032140} (\bibinfo {year} {2016})}\BibitemShut
  {NoStop}%
\bibitem [{\citenamefont {Cramer}\ \emph {et~al.}(2010)\citenamefont {Cramer},
  \citenamefont {Plenio}, \citenamefont {Flammia}, \citenamefont {Somma},
  \citenamefont {Gross}, \citenamefont {Bartlett}, \citenamefont
  {Landon-Cardinal}, \citenamefont {Poulin},\ and\ \citenamefont
  {Liu}}]{efficient_Cramer_2010}%
  \BibitemOpen
  \bibfield  {author} {\bibinfo {author} {\bibfnamefont {M.}~\bibnamefont
  {Cramer}}, \bibinfo {author} {\bibfnamefont {M.~B.}\ \bibnamefont {Plenio}},
  \bibinfo {author} {\bibfnamefont {S.~T.}\ \bibnamefont {Flammia}}, \bibinfo
  {author} {\bibfnamefont {R.}~\bibnamefont {Somma}}, \bibinfo {author}
  {\bibfnamefont {D.}~\bibnamefont {Gross}}, \bibinfo {author} {\bibfnamefont
  {S.~D.}\ \bibnamefont {Bartlett}}, \bibinfo {author} {\bibfnamefont
  {O.}~\bibnamefont {Landon-Cardinal}}, \bibinfo {author} {\bibfnamefont
  {D.}~\bibnamefont {Poulin}},\ and\ \bibinfo {author} {\bibfnamefont {Y.-K.}\
  \bibnamefont {Liu}},\ }\bibfield  {title} {\bibinfo {title} {Efficient
  quantum state tomography},\ }\href {https://doi.org/10.1038/ncomms1147}
  {\bibfield  {journal} {\bibinfo  {journal} {Nature Communications}\ }\textbf
  {\bibinfo {volume} {1}},\ \bibinfo {pages} {1} (\bibinfo {year}
  {2010})}\BibitemShut {NoStop}%
\bibitem [{\citenamefont {Qi}\ \emph {et~al.}(2013)\citenamefont {Qi},
  \citenamefont {Hou}, \citenamefont {Li}, \citenamefont {Dong}, \citenamefont
  {Xiang},\ and\ \citenamefont {Guo}}]{LRE_qi_2013}%
  \BibitemOpen
  \bibfield  {author} {\bibinfo {author} {\bibfnamefont {B.}~\bibnamefont
  {Qi}}, \bibinfo {author} {\bibfnamefont {Z.}~\bibnamefont {Hou}}, \bibinfo
  {author} {\bibfnamefont {L.}~\bibnamefont {Li}}, \bibinfo {author}
  {\bibfnamefont {D.}~\bibnamefont {Dong}}, \bibinfo {author} {\bibfnamefont
  {G.}~\bibnamefont {Xiang}},\ and\ \bibinfo {author} {\bibfnamefont
  {G.}~\bibnamefont {Guo}},\ }\bibfield  {title} {\bibinfo {title} {Quantum
  {State} {Tomography} via {Linear} {Regression} {Estimation}},\ }\href
  {https://doi.org/10.1038/srep03496} {\bibfield  {journal} {\bibinfo
  {journal} {Scientific Reports}\ }\textbf {\bibinfo {volume} {3}},\ \bibinfo
  {pages} {3496} (\bibinfo {year} {2013})}\BibitemShut {NoStop}%
\bibitem [{\citenamefont {Díaz}\ \emph {et~al.}(2015)\citenamefont {Díaz},
  \citenamefont {Sainz},\ and\ \citenamefont {Klimov}}]{QSTnob_Diaz_2015}%
  \BibitemOpen
  \bibfield  {author} {\bibinfo {author} {\bibfnamefont {J.~J.}\ \bibnamefont
  {Díaz}}, \bibinfo {author} {\bibfnamefont {I.}~\bibnamefont {Sainz}},\ and\
  \bibinfo {author} {\bibfnamefont {A.~B.}\ \bibnamefont {Klimov}},\ }\bibfield
   {title} {\bibinfo {title} {Quantum tomography via nonorthogonal basis and
  weak values},\ }\href {https://doi.org/10.1103/PhysRevA.91.062127} {\bibfield
   {journal} {\bibinfo  {journal} {Physical Review A}\ }\textbf {\bibinfo
  {volume} {91}},\ \bibinfo {pages} {062127} (\bibinfo {year}
  {2015})}\BibitemShut {NoStop}%
\bibitem [{\citenamefont
  {Di~Lorenzo}(2013{\natexlab{a}})}]{ss-seq_diLorenzo_2013}%
  \BibitemOpen
  \bibfield  {author} {\bibinfo {author} {\bibfnamefont {A.}~\bibnamefont
  {Di~Lorenzo}},\ }\bibfield  {title} {\bibinfo {title} {Quantum state
  tomography from a sequential measurement of two variables in a single
  setup},\ }\href {https://doi.org/10.1103/PhysRevA.88.042114} {\bibfield
  {journal} {\bibinfo  {journal} {Phys. Rev. A}\ }\textbf {\bibinfo {volume}
  {88}},\ \bibinfo {pages} {042114} (\bibinfo {year}
  {2013}{\natexlab{a}})}\BibitemShut {NoStop}%
\bibitem [{\citenamefont {Lin}\ and\ \citenamefont
  {Jovanovic}(2016)}]{ss-pulse_lin_2016}%
  \BibitemOpen
  \bibfield  {author} {\bibinfo {author} {\bibfnamefont {M.-W.}\ \bibnamefont
  {Lin}}\ and\ \bibinfo {author} {\bibfnamefont {I.}~\bibnamefont
  {Jovanovic}},\ }\bibfield  {title} {\bibinfo {title} {Single-{Shot}
  {Measurement} of {Temporally}-{Dependent} {Polarization} {State} of
  {Femtosecond} {Pulses} by {Angle}-{Multiplexed} {Spectral}-{Spatial}
  {Interferometry}},\ }\href {https://doi.org/10.1038/srep32839} {\bibfield
  {journal} {\bibinfo  {journal} {Scientific Reports}\ }\textbf {\bibinfo
  {volume} {6}},\ \bibinfo {pages} {32839} (\bibinfo {year}
  {2016})}\BibitemShut {NoStop}%
\bibitem [{\citenamefont {Zhu}\ \emph {et~al.}(2019)\citenamefont {Zhu},
  \citenamefont {Hay}, \citenamefont {Zhou}, \citenamefont {Fyffe},
  \citenamefont {Kantor}, \citenamefont {Agarwal}, \citenamefont {Boyd},\ and\
  \citenamefont {Shi}}]{ss-trans_zhu_2019}%
  \BibitemOpen
  \bibfield  {author} {\bibinfo {author} {\bibfnamefont {Z.}~\bibnamefont
  {Zhu}}, \bibinfo {author} {\bibfnamefont {D.}~\bibnamefont {Hay}}, \bibinfo
  {author} {\bibfnamefont {Y.}~\bibnamefont {Zhou}}, \bibinfo {author}
  {\bibfnamefont {A.}~\bibnamefont {Fyffe}}, \bibinfo {author} {\bibfnamefont
  {B.}~\bibnamefont {Kantor}}, \bibinfo {author} {\bibfnamefont {G.~S.}\
  \bibnamefont {Agarwal}}, \bibinfo {author} {\bibfnamefont {R.~W.}\
  \bibnamefont {Boyd}},\ and\ \bibinfo {author} {\bibfnamefont
  {Z.}~\bibnamefont {Shi}},\ }\bibfield  {title} {\bibinfo {title}
  {Single-{Shot} {Direct} {Tomography} of the {Complete} {Transverse}
  {Amplitude}, {Phase}, and {Polarization} {Structure} of a {Light} {Field}},\
  }\href {https://doi.org/10.1103/PhysRevApplied.12.034036} {\bibfield
  {journal} {\bibinfo  {journal} {Physical Review Applied}\ }\textbf {\bibinfo
  {volume} {12}},\ \bibinfo {pages} {034036} (\bibinfo {year}
  {2019})}\BibitemShut {NoStop}%
\bibitem [{\citenamefont {B~S}\ \emph {et~al.}(2020)\citenamefont {B~S},
  \citenamefont {Pal}, \citenamefont {Mukherjee}, \citenamefont {Mishra},
  \citenamefont {Nandy},\ and\ \citenamefont {Ghosh}}]{ss-geo_athira_2019}%
  \BibitemOpen
  \bibfield  {author} {\bibinfo {author} {\bibfnamefont {A.}~\bibnamefont
  {B~S}}, \bibinfo {author} {\bibfnamefont {M.}~\bibnamefont {Pal}}, \bibinfo
  {author} {\bibfnamefont {S.}~\bibnamefont {Mukherjee}}, \bibinfo {author}
  {\bibfnamefont {J.}~\bibnamefont {Mishra}}, \bibinfo {author} {\bibfnamefont
  {D.}~\bibnamefont {Nandy}},\ and\ \bibinfo {author} {\bibfnamefont
  {N.}~\bibnamefont {Ghosh}},\ }\bibfield  {title} {\bibinfo {title}
  {Single-shot measurement of the space-varying polarization state of light
  through interferometric quantification of the geometric phase},\ }\href
  {https://doi.org/10.1103/PhysRevA.101.013836} {\bibfield  {journal} {\bibinfo
   {journal} {Phys. Rev. A}\ }\textbf {\bibinfo {volume} {101}},\ \bibinfo
  {pages} {013836} (\bibinfo {year} {2020})}\BibitemShut {NoStop}%
\bibitem [{\citenamefont
  {Di~Lorenzo}(2013{\natexlab{b}})}]{alt-seq_diLorenzo_2013}%
  \BibitemOpen
  \bibfield  {author} {\bibinfo {author} {\bibfnamefont {A.}~\bibnamefont
  {Di~Lorenzo}},\ }\bibfield  {title} {\bibinfo {title} {Sequential measurement
  of conjugate variables as an alternative quantum state tomography},\ }\href
  {https://doi.org/10.1103/PhysRevLett.110.010404} {\bibfield  {journal}
  {\bibinfo  {journal} {Phys. Rev. Lett.}\ }\textbf {\bibinfo {volume} {110}},\
  \bibinfo {pages} {010404} (\bibinfo {year} {2013}{\natexlab{b}})}\BibitemShut
  {NoStop}%
\bibitem [{\citenamefont {Wu}(2013)}]{weakQST_wu_2013}%
  \BibitemOpen
  \bibfield  {author} {\bibinfo {author} {\bibfnamefont {S.}~\bibnamefont
  {Wu}},\ }\bibfield  {title} {\bibinfo {title} {State tomography via weak
  measurements},\ }\href {https://doi.org/10.1038/srep01193} {\bibfield
  {journal} {\bibinfo  {journal} {Scientific Reports}\ }\textbf {\bibinfo
  {volume} {3}},\ \bibinfo {pages} {1193} (\bibinfo {year} {2013})}\BibitemShut
  {NoStop}%
\bibitem [{\citenamefont {Chen}\ \emph {et~al.}(2018)\citenamefont {Chen},
  \citenamefont {Dai}, \citenamefont {Yang},\ and\ \citenamefont
  {Zhang}}]{alt-weak_Chen_2018}%
  \BibitemOpen
  \bibfield  {author} {\bibinfo {author} {\bibfnamefont {X.}~\bibnamefont
  {Chen}}, \bibinfo {author} {\bibfnamefont {H.-Y.}\ \bibnamefont {Dai}},
  \bibinfo {author} {\bibfnamefont {L.}~\bibnamefont {Yang}},\ and\ \bibinfo
  {author} {\bibfnamefont {M.}~\bibnamefont {Zhang}},\ }\bibfield  {title}
  {\bibinfo {title} {Alternative method of quantum state tomography toward a
  typical target via a weak-value measurement},\ }\href
  {https://doi.org/10.1103/PhysRevA.97.032120} {\bibfield  {journal} {\bibinfo
  {journal} {Phys. Rev. A}\ }\textbf {\bibinfo {volume} {97}},\ \bibinfo
  {pages} {032120} (\bibinfo {year} {2018})}\BibitemShut {NoStop}%
\bibitem [{\citenamefont {Shojaee}\ \emph {et~al.}(2018)\citenamefont
  {Shojaee}, \citenamefont {Jackson}, \citenamefont {Riofr\'{\i}o},
  \citenamefont {Kalev},\ and\ \citenamefont {Deutsch}}]{alt-opt_Shojaee_2018}%
  \BibitemOpen
  \bibfield  {author} {\bibinfo {author} {\bibfnamefont {E.}~\bibnamefont
  {Shojaee}}, \bibinfo {author} {\bibfnamefont {C.~S.}\ \bibnamefont
  {Jackson}}, \bibinfo {author} {\bibfnamefont {C.~A.}\ \bibnamefont
  {Riofr\'{\i}o}}, \bibinfo {author} {\bibfnamefont {A.}~\bibnamefont
  {Kalev}},\ and\ \bibinfo {author} {\bibfnamefont {I.~H.}\ \bibnamefont
  {Deutsch}},\ }\bibfield  {title} {\bibinfo {title} {Optimal pure-state qubit
  tomography via sequential weak measurements},\ }\href
  {https://doi.org/10.1103/PhysRevLett.121.130404} {\bibfield  {journal}
  {\bibinfo  {journal} {Phys. Rev. Lett.}\ }\textbf {\bibinfo {volume} {121}},\
  \bibinfo {pages} {130404} (\bibinfo {year} {2018})}\BibitemShut {NoStop}%
\bibitem [{\citenamefont {Aharonov}\ \emph {et~al.}(1988)\citenamefont
  {Aharonov}, \citenamefont {Albert},\ and\ \citenamefont
  {Vaidman}}]{wm_Aharonov_1988}%
  \BibitemOpen
  \bibfield  {author} {\bibinfo {author} {\bibfnamefont {Y.}~\bibnamefont
  {Aharonov}}, \bibinfo {author} {\bibfnamefont {D.~Z.}\ \bibnamefont
  {Albert}},\ and\ \bibinfo {author} {\bibfnamefont {L.}~\bibnamefont
  {Vaidman}},\ }\bibfield  {title} {\bibinfo {title} {How the result of a
  measurement of a component of the spin of a spin-1/2 particle can turn out to
  be 100},\ }\href {https://doi.org/10.1103/PhysRevLett.60.1351} {\bibfield
  {journal} {\bibinfo  {journal} {Phys. Rev. Lett.}\ }\textbf {\bibinfo
  {volume} {60}},\ \bibinfo {pages} {1351} (\bibinfo {year}
  {1988})}\BibitemShut {NoStop}%
\bibitem [{\citenamefont {Duck}\ \emph {et~al.}(1989)\citenamefont {Duck},
  \citenamefont {Stevenson},\ and\ \citenamefont {Sudarshan}}]{wm_Duck_1989}%
  \BibitemOpen
  \bibfield  {author} {\bibinfo {author} {\bibfnamefont {I.~M.}\ \bibnamefont
  {Duck}}, \bibinfo {author} {\bibfnamefont {P.~M.}\ \bibnamefont
  {Stevenson}},\ and\ \bibinfo {author} {\bibfnamefont {E.~C.~G.}\ \bibnamefont
  {Sudarshan}},\ }\bibfield  {title} {\bibinfo {title} {The sense in which a
  "weak measurement" of a spin-$1/2$ particle's spin component yields a value
  100},\ }\href {https://doi.org/10.1103/PhysRevD.40.2112} {\bibfield
  {journal} {\bibinfo  {journal} {Phys. Rev. D}\ }\textbf {\bibinfo {volume}
  {40}},\ \bibinfo {pages} {2112} (\bibinfo {year} {1989})}\BibitemShut
  {NoStop}%
\bibitem [{\citenamefont {Lundeen}\ \emph {et~al.}(2011)\citenamefont
  {Lundeen}, \citenamefont {Sutherland}, \citenamefont {Patel}, \citenamefont
  {Stewart},\ and\ \citenamefont {Bamber}}]{dm-wf_Lundeen_2011}%
  \BibitemOpen
  \bibfield  {author} {\bibinfo {author} {\bibfnamefont {J.~S.}\ \bibnamefont
  {Lundeen}}, \bibinfo {author} {\bibfnamefont {B.}~\bibnamefont {Sutherland}},
  \bibinfo {author} {\bibfnamefont {A.}~\bibnamefont {Patel}}, \bibinfo
  {author} {\bibfnamefont {C.}~\bibnamefont {Stewart}},\ and\ \bibinfo {author}
  {\bibfnamefont {C.}~\bibnamefont {Bamber}},\ }\bibfield  {title} {\bibinfo
  {title} {Direct measurement of the quantum wavefunction},\ }\href
  {https://doi.org/10.1038/nature10120} {\bibfield  {journal} {\bibinfo
  {journal} {Nature}\ }\textbf {\bibinfo {volume} {474}},\ \bibinfo {pages}
  {188} (\bibinfo {year} {2011})}\BibitemShut {NoStop}%
\bibitem [{\citenamefont {Lundeen}\ and\ \citenamefont
  {Bamber}(2012)}]{dm-procedure_Lundeen_2012}%
  \BibitemOpen
  \bibfield  {author} {\bibinfo {author} {\bibfnamefont {J.~S.}\ \bibnamefont
  {Lundeen}}\ and\ \bibinfo {author} {\bibfnamefont {C.}~\bibnamefont
  {Bamber}},\ }\bibfield  {title} {\bibinfo {title} {Procedure for {Direct}
  {Measurement} of {General} {Quantum} {States} {Using} {Weak} {Measurement}},\
  }\href {https://doi.org/10.1103/PhysRevLett.108.070402} {\bibfield  {journal}
  {\bibinfo  {journal} {Physical Review Letters}\ }\textbf {\bibinfo {volume}
  {108}},\ \bibinfo {pages} {070402} (\bibinfo {year} {2012})}\BibitemShut
  {NoStop}%
\bibitem [{\citenamefont {Thekkadath}\ \emph {et~al.}(2016)\citenamefont
  {Thekkadath}, \citenamefont {Giner}, \citenamefont {Chalich}, \citenamefont
  {Horton}, \citenamefont {Banker},\ and\ \citenamefont
  {Lundeen}}]{dm-dm_thekkadath_2016}%
  \BibitemOpen
  \bibfield  {author} {\bibinfo {author} {\bibfnamefont {G.}~\bibnamefont
  {Thekkadath}}, \bibinfo {author} {\bibfnamefont {L.}~\bibnamefont {Giner}},
  \bibinfo {author} {\bibfnamefont {Y.}~\bibnamefont {Chalich}}, \bibinfo
  {author} {\bibfnamefont {M.}~\bibnamefont {Horton}}, \bibinfo {author}
  {\bibfnamefont {J.}~\bibnamefont {Banker}},\ and\ \bibinfo {author}
  {\bibfnamefont {J.}~\bibnamefont {Lundeen}},\ }\bibfield  {title} {\bibinfo
  {title} {Direct {Measurement} of the {Density} {Matrix} of a {Quantum}
  {System}},\ }\href {https://doi.org/10.1103/PhysRevLett.117.120401}
  {\bibfield  {journal} {\bibinfo  {journal} {Physical Review Letters}\
  }\textbf {\bibinfo {volume} {117}},\ \bibinfo {pages} {120401} (\bibinfo
  {year} {2016})}\BibitemShut {NoStop}%
\bibitem [{\citenamefont {Salvail}\ \emph {et~al.}(2013)\citenamefont
  {Salvail}, \citenamefont {Agnew}, \citenamefont {Johnson}, \citenamefont
  {Bolduc}, \citenamefont {Leach},\ and\ \citenamefont
  {Boyd}}]{dm-pol_salvail_2013}%
  \BibitemOpen
  \bibfield  {author} {\bibinfo {author} {\bibfnamefont {J.~Z.}\ \bibnamefont
  {Salvail}}, \bibinfo {author} {\bibfnamefont {M.}~\bibnamefont {Agnew}},
  \bibinfo {author} {\bibfnamefont {A.~S.}\ \bibnamefont {Johnson}}, \bibinfo
  {author} {\bibfnamefont {E.}~\bibnamefont {Bolduc}}, \bibinfo {author}
  {\bibfnamefont {J.}~\bibnamefont {Leach}},\ and\ \bibinfo {author}
  {\bibfnamefont {R.~W.}\ \bibnamefont {Boyd}},\ }\bibfield  {title} {\bibinfo
  {title} {Full characterization of polarization states of light via direct
  measurement},\ }\href {https://doi.org/10.1038/nphoton.2013.24} {\bibfield
  {journal} {\bibinfo  {journal} {Nature Photonics}\ }\textbf {\bibinfo
  {volume} {7}},\ \bibinfo {pages} {316} (\bibinfo {year} {2013})}\BibitemShut
  {NoStop}%
\bibitem [{\citenamefont {Malik}\ \emph {et~al.}(2014)\citenamefont {Malik},
  \citenamefont {Mirhosseini}, \citenamefont {Lavery}, \citenamefont {Leach},
  \citenamefont {Padgett},\ and\ \citenamefont {Boyd}}]{dm27_Malik_2014}%
  \BibitemOpen
  \bibfield  {author} {\bibinfo {author} {\bibfnamefont {M.}~\bibnamefont
  {Malik}}, \bibinfo {author} {\bibfnamefont {M.}~\bibnamefont {Mirhosseini}},
  \bibinfo {author} {\bibfnamefont {M.~P.~J.}\ \bibnamefont {Lavery}}, \bibinfo
  {author} {\bibfnamefont {J.}~\bibnamefont {Leach}}, \bibinfo {author}
  {\bibfnamefont {M.~J.}\ \bibnamefont {Padgett}},\ and\ \bibinfo {author}
  {\bibfnamefont {R.~W.}\ \bibnamefont {Boyd}},\ }\bibfield  {title} {\bibinfo
  {title} {Direct measurement of a 27-dimensional orbital-angular-momentum
  state vector},\ }\href {https://doi.org/10.1038/ncomms4115} {\bibfield
  {journal} {\bibinfo  {journal} {Nature Communications}\ }\textbf {\bibinfo
  {volume} {5}},\ \bibinfo {pages} {3115} (\bibinfo {year} {2014})}\BibitemShut
  {NoStop}%
\bibitem [{\citenamefont {Mirhosseini}\ \emph {et~al.}(2014)\citenamefont
  {Mirhosseini}, \citenamefont {Magaña-Loaiza}, \citenamefont
  {Hashemi~Rafsanjani},\ and\ \citenamefont
  {Boyd}}]{dm-compressive_mirhosseini_2014}%
  \BibitemOpen
  \bibfield  {author} {\bibinfo {author} {\bibfnamefont {M.}~\bibnamefont
  {Mirhosseini}}, \bibinfo {author} {\bibfnamefont {O.~S.}\ \bibnamefont
  {Magaña-Loaiza}}, \bibinfo {author} {\bibfnamefont {S.~M.}\ \bibnamefont
  {Hashemi~Rafsanjani}},\ and\ \bibinfo {author} {\bibfnamefont {R.~W.}\
  \bibnamefont {Boyd}},\ }\bibfield  {title} {\bibinfo {title} {Compressive
  {Direct} {Measurement} of the {Quantum} {Wave} {Function}},\ }\href
  {https://doi.org/10.1103/PhysRevLett.113.090402} {\bibfield  {journal}
  {\bibinfo  {journal} {Physical Review Letters}\ }\textbf {\bibinfo {volume}
  {113}},\ \bibinfo {pages} {090402} (\bibinfo {year} {2014})}\BibitemShut
  {NoStop}%
\bibitem [{\citenamefont {{Zhimin Shi}}\ \emph {et~al.}(2016)\citenamefont
  {{Zhimin Shi}}, \citenamefont {{Mirhosseini}}, \citenamefont {{Margiewicz}},
  \citenamefont {{Malik}}, \citenamefont {{Rivera}}, \citenamefont {{Ziyi
  Zhu}},\ and\ \citenamefont {{Boyd}}}]{dm1M_Zhimin_2016}%
  \BibitemOpen
  \bibfield  {author} {\bibinfo {author} {\bibnamefont {{Zhimin Shi}}},
  \bibinfo {author} {\bibfnamefont {M.}~\bibnamefont {{Mirhosseini}}}, \bibinfo
  {author} {\bibfnamefont {J.}~\bibnamefont {{Margiewicz}}}, \bibinfo {author}
  {\bibfnamefont {M.}~\bibnamefont {{Malik}}}, \bibinfo {author} {\bibfnamefont
  {F.}~\bibnamefont {{Rivera}}}, \bibinfo {author} {\bibnamefont {{Ziyi
  Zhu}}},\ and\ \bibinfo {author} {\bibfnamefont {R.~W.}\ \bibnamefont
  {{Boyd}}},\ }\bibfield  {title} {\bibinfo {title} {Direct measurement of an
  one-million-dimensional photonic state},\ }in\ \href
  {https://doi.org/10.1109/PIERS.2016.7734289} {\emph {\bibinfo {booktitle}
  {2016 Progress in Electromagnetic Research Symposium (PIERS)}}}\ (\bibinfo
  {year} {2016})\ pp.\ \bibinfo {pages} {187--187}\BibitemShut {NoStop}%
\bibitem [{\citenamefont {Nirala}\ \emph {et~al.}(2019)\citenamefont {Nirala},
  \citenamefont {Sahoo}, \citenamefont {Pati},\ and\ \citenamefont
  {Sinha}}]{non_hermitian_expt_Nirala_2019}%
  \BibitemOpen
  \bibfield  {author} {\bibinfo {author} {\bibfnamefont {G.}~\bibnamefont
  {Nirala}}, \bibinfo {author} {\bibfnamefont {S.~N.}\ \bibnamefont {Sahoo}},
  \bibinfo {author} {\bibfnamefont {A.~K.}\ \bibnamefont {Pati}},\ and\
  \bibinfo {author} {\bibfnamefont {U.}~\bibnamefont {Sinha}},\ }\bibfield
  {title} {\bibinfo {title} {Measuring average of non-hermitian operator with
  weak value in a mach-zehnder interferometer},\ }\href
  {https://doi.org/10.1103/PhysRevA.99.022111} {\bibfield  {journal} {\bibinfo
  {journal} {Phys. Rev. A}\ }\textbf {\bibinfo {volume} {99}},\ \bibinfo
  {pages} {022111} (\bibinfo {year} {2019})}\BibitemShut {NoStop}%
\bibitem [{\citenamefont {Abbott}\ \emph {et~al.}(2019)\citenamefont {Abbott},
  \citenamefont {Silva}, \citenamefont {Wechs}, \citenamefont {Brunner},\ and\
  \citenamefont {Branciard}}]{anomalousweakvalues_Abbott_2019}%
  \BibitemOpen
  \bibfield  {author} {\bibinfo {author} {\bibfnamefont {A.~A.}\ \bibnamefont
  {Abbott}}, \bibinfo {author} {\bibfnamefont {R.}~\bibnamefont {Silva}},
  \bibinfo {author} {\bibfnamefont {J.}~\bibnamefont {Wechs}}, \bibinfo
  {author} {\bibfnamefont {N.}~\bibnamefont {Brunner}},\ and\ \bibinfo {author}
  {\bibfnamefont {C.}~\bibnamefont {Branciard}},\ }\bibfield  {title} {\bibinfo
  {title} {Anomalous {W}eak {V}alues {W}ithout {P}ost-{S}election},\ }\href
  {https://doi.org/10.22331/q-2019-10-14-194} {\bibfield  {journal} {\bibinfo
  {journal} {{Quantum}}\ }\textbf {\bibinfo {volume} {3}},\ \bibinfo {pages}
  {194} (\bibinfo {year} {2019})}\BibitemShut {NoStop}%
\bibitem [{\citenamefont {Ogawa}\ \emph {et~al.}(2019)\citenamefont {Ogawa},
  \citenamefont {Yasuhiko}, \citenamefont {Kobayashi}, \citenamefont
  {Nakanishi},\ and\ \citenamefont {Tomita}}]{wv-framework_Ogawa_2019}%
  \BibitemOpen
  \bibfield  {author} {\bibinfo {author} {\bibfnamefont {K.}~\bibnamefont
  {Ogawa}}, \bibinfo {author} {\bibfnamefont {O.}~\bibnamefont {Yasuhiko}},
  \bibinfo {author} {\bibfnamefont {H.}~\bibnamefont {Kobayashi}}, \bibinfo
  {author} {\bibfnamefont {T.}~\bibnamefont {Nakanishi}},\ and\ \bibinfo
  {author} {\bibfnamefont {A.}~\bibnamefont {Tomita}},\ }\bibfield  {title}
  {\bibinfo {title} {A framework for measuring weak values without weak
  interactions and its diagrammatic representation},\ }\href
  {https://doi.org/10.1088/1367-2630/ab0773} {\bibfield  {journal} {\bibinfo
  {journal} {New Journal of Physics}\ }\textbf {\bibinfo {volume} {21}},\
  \bibinfo {pages} {043013} (\bibinfo {year} {2019})}\BibitemShut {NoStop}%
\bibitem [{\citenamefont {Pati}\ \emph {et~al.}(2015)\citenamefont {Pati},
  \citenamefont {Singh},\ and\ \citenamefont
  {Sinha}}]{non_hermitian_theory_Pati_2015}%
  \BibitemOpen
  \bibfield  {author} {\bibinfo {author} {\bibfnamefont {A.~K.}\ \bibnamefont
  {Pati}}, \bibinfo {author} {\bibfnamefont {U.}~\bibnamefont {Singh}},\ and\
  \bibinfo {author} {\bibfnamefont {U.}~\bibnamefont {Sinha}},\ }\bibfield
  {title} {\bibinfo {title} {Measuring non-hermitian operators via weak
  values},\ }\href {https://doi.org/10.1103/PhysRevA.92.052120} {\bibfield
  {journal} {\bibinfo  {journal} {Phys. Rev. A}\ }\textbf {\bibinfo {volume}
  {92}},\ \bibinfo {pages} {052120} (\bibinfo {year} {2015})}\BibitemShut
  {NoStop}%
\bibitem [{\citenamefont {Bolduc}\ \emph {et~al.}(2016)\citenamefont {Bolduc},
  \citenamefont {Gariepy},\ and\ \citenamefont
  {Leach}}]{directNonHermitian_bolduc_2016}%
  \BibitemOpen
  \bibfield  {author} {\bibinfo {author} {\bibfnamefont {E.}~\bibnamefont
  {Bolduc}}, \bibinfo {author} {\bibfnamefont {G.}~\bibnamefont {Gariepy}},\
  and\ \bibinfo {author} {\bibfnamefont {J.}~\bibnamefont {Leach}},\ }\bibfield
   {title} {\bibinfo {title} {Direct measurement of large-scale quantum states
  via expectation values of non-{Hermitian} matrices},\ }\href
  {https://doi.org/10.1038/ncomms10439} {\bibfield  {journal} {\bibinfo
  {journal} {Nature Communications}\ }\textbf {\bibinfo {volume} {7}},\
  \bibinfo {pages} {10439} (\bibinfo {year} {2016})}\BibitemShut {NoStop}%
\bibitem [{Note1()}]{Note1}%
  \BibitemOpen
  \bibinfo {note} {Here, by single-shot, we mean that the measurement settings
  do not change during the course of data acquisition}\BibitemShut {NoStop}%
\bibitem [{\citenamefont {Flamini}\ \emph {et~al.}(2018)\citenamefont
  {Flamini}, \citenamefont {Spagnolo},\ and\ \citenamefont
  {Sciarrino}}]{photonicReview_Flamini_2018}%
  \BibitemOpen
  \bibfield  {author} {\bibinfo {author} {\bibfnamefont {F.}~\bibnamefont
  {Flamini}}, \bibinfo {author} {\bibfnamefont {N.}~\bibnamefont {Spagnolo}},\
  and\ \bibinfo {author} {\bibfnamefont {F.}~\bibnamefont {Sciarrino}},\
  }\bibfield  {title} {\bibinfo {title} {Photonic quantum information
  processing: a review},\ }\href {https://doi.org/10.1088/1361-6633/aad5b2}
  {\bibfield  {journal} {\bibinfo  {journal} {Reports on Progress in Physics}\
  }\textbf {\bibinfo {volume} {82}},\ \bibinfo {pages} {016001} (\bibinfo
  {year} {2018})}\BibitemShut {NoStop}%
\bibitem [{\citenamefont {Slussarenko}\ and\ \citenamefont
  {Pryde}(2019)}]{photonicReview_slussarenko_2019}%
  \BibitemOpen
  \bibfield  {author} {\bibinfo {author} {\bibfnamefont {S.}~\bibnamefont
  {Slussarenko}}\ and\ \bibinfo {author} {\bibfnamefont {G.~J.}\ \bibnamefont
  {Pryde}},\ }\bibfield  {title} {\bibinfo {title} {Photonic quantum
  information processing: {A} concise review},\ }\href
  {https://doi.org/10.1063/1.5115814} {\bibfield  {journal} {\bibinfo
  {journal} {Applied Physics Reviews}\ }\textbf {\bibinfo {volume} {6}},\
  \bibinfo {pages} {041303} (\bibinfo {year} {2019})}\BibitemShut {NoStop}%
\bibitem [{\citenamefont {Mičuda}\ \emph {et~al.}(2014)\citenamefont
  {Mičuda}, \citenamefont {Doláková}, \citenamefont {Straka}, \citenamefont
  {Miková}, \citenamefont {Dušek}, \citenamefont {Fiurášek},\ and\
  \citenamefont {Ježek}}]{displacedSagnac_Micuda_2014}%
  \BibitemOpen
  \bibfield  {author} {\bibinfo {author} {\bibfnamefont {M.}~\bibnamefont
  {Mičuda}}, \bibinfo {author} {\bibfnamefont {E.}~\bibnamefont {Doláková}},
  \bibinfo {author} {\bibfnamefont {I.}~\bibnamefont {Straka}}, \bibinfo
  {author} {\bibfnamefont {M.}~\bibnamefont {Miková}}, \bibinfo {author}
  {\bibfnamefont {M.}~\bibnamefont {Dušek}}, \bibinfo {author} {\bibfnamefont
  {J.}~\bibnamefont {Fiurášek}},\ and\ \bibinfo {author} {\bibfnamefont
  {M.}~\bibnamefont {Ježek}},\ }\bibfield  {title} {\bibinfo {title} {Highly
  stable polarization independent mach-zehnder interferometer},\ }\href
  {https://doi.org/10.1063/1.4891702} {\bibfield  {journal} {\bibinfo
  {journal} {Review of Scientific Instruments}\ }\textbf {\bibinfo {volume}
  {85}},\ \bibinfo {pages} {083103} (\bibinfo {year} {2014})},\ \Eprint
  {https://arxiv.org/abs/https://doi.org/10.1063/1.4891702}
  {https://doi.org/10.1063/1.4891702} \BibitemShut {NoStop}%
\bibitem [{Note2()}]{Note2}%
  \BibitemOpen
  \bibinfo {note} {See Supplementary for derivation}\BibitemShut {NoStop}%
\bibitem [{\citenamefont {Bennett}\ \emph {et~al.}(1996)\citenamefont
  {Bennett}, \citenamefont {Bernstein}, \citenamefont {Popescu},\ and\
  \citenamefont {Schumacher}}]{entanglementEntropy_Bennet_1996}%
  \BibitemOpen
  \bibfield  {author} {\bibinfo {author} {\bibfnamefont {C.~H.}\ \bibnamefont
  {Bennett}}, \bibinfo {author} {\bibfnamefont {H.~J.}\ \bibnamefont
  {Bernstein}}, \bibinfo {author} {\bibfnamefont {S.}~\bibnamefont {Popescu}},\
  and\ \bibinfo {author} {\bibfnamefont {B.}~\bibnamefont {Schumacher}},\
  }\bibfield  {title} {\bibinfo {title} {Concentrating partial entanglement by
  local operations},\ }\href {https://doi.org/10.1103/PhysRevA.53.2046}
  {\bibfield  {journal} {\bibinfo  {journal} {Phys. Rev. A}\ }\textbf {\bibinfo
  {volume} {53}},\ \bibinfo {pages} {2046} (\bibinfo {year}
  {1996})}\BibitemShut {NoStop}%
\bibitem [{\citenamefont {Horodecki}\ \emph {et~al.}(2009)\citenamefont
  {Horodecki}, \citenamefont {Horodecki}, \citenamefont {Horodecki},\ and\
  \citenamefont {Horodecki}}]{entanglementEntropy_Horodeki_2009}%
  \BibitemOpen
  \bibfield  {author} {\bibinfo {author} {\bibfnamefont {R.}~\bibnamefont
  {Horodecki}}, \bibinfo {author} {\bibfnamefont {P.}~\bibnamefont
  {Horodecki}}, \bibinfo {author} {\bibfnamefont {M.}~\bibnamefont
  {Horodecki}},\ and\ \bibinfo {author} {\bibfnamefont {K.}~\bibnamefont
  {Horodecki}},\ }\bibfield  {title} {\bibinfo {title} {Quantum entanglement},\
  }\href {https://doi.org/10.1103/RevModPhys.81.865} {\bibfield  {journal}
  {\bibinfo  {journal} {Rev. Mod. Phys.}\ }\textbf {\bibinfo {volume} {81}},\
  \bibinfo {pages} {865} (\bibinfo {year} {2009})}\BibitemShut {NoStop}%
\bibitem [{\citenamefont {{Sorkin}}(1983)}]{entanglementEntropy_Sorkin_1983}%
  \BibitemOpen
  \bibfield  {author} {\bibinfo {author} {\bibfnamefont {R.~D.}\ \bibnamefont
  {{Sorkin}}},\ }\bibfield  {title} {\bibinfo {title} {{On the Entropy of the
  Vacuum Outside a Horizon}},\ }in\ \href@noop {} {\emph {\bibinfo {booktitle}
  {General Relativity and Gravitation, Volume 1}}},\ Vol.~\bibinfo {volume}
  {1},\ \bibinfo {editor} {edited by\ \bibinfo {editor} {\bibfnamefont
  {B.}~\bibnamefont {{Bertotti}}}, \bibinfo {editor} {\bibfnamefont
  {F.}~\bibnamefont {{de Felice}}},\ and\ \bibinfo {editor} {\bibfnamefont
  {A.}~\bibnamefont {{Pascolini}}}}\ (\bibinfo {year} {1983})\ p.\ \bibinfo
  {pages} {734}\BibitemShut {NoStop}%
\bibitem [{Note3()}]{Note3}%
  \BibitemOpen
  \bibinfo {note} {The state reconstruction of bipartite qubit pure states is
  detailed in the \protect \textit {Supplementary Material Sec.}
  $XIV$}\BibitemShut {NoStop}%
\bibitem [{Note4()}]{Note4}%
  \BibitemOpen
  \bibinfo {note} {See supplementary for a comparison between interferometers
  for QSI, which includes Ref. [51]}\BibitemShut {NoStop}%
\bibitem [{\citenamefont {Fisher}(1993)}]{circstats_fisher_1993}%
  \BibitemOpen
  \bibfield  {author} {\bibinfo {author} {\bibfnamefont {N.~I.}\ \bibnamefont
  {Fisher}},\ }\href {https://doi.org/10.1017/CBO9780511564345} {\emph
  {\bibinfo {title} {Statistical Analysis of Circular Data}}}\ (\bibinfo
  {publisher} {Cambridge University Press},\ \bibinfo {year}
  {1993})\BibitemShut {NoStop}%
\bibitem [{Note5()}]{Note5}%
  \BibitemOpen
  \bibinfo {note} {The details of how $\mu $ affects the fidelity and other
  methods to obtain $\mu $ from experiment that makes $\rho $ physical is
  discussed in \protect \textit {Supplementary Material Sec VIII}}\BibitemShut
  {NoStop}%
\bibitem [{\citenamefont {Blumenson}(1960)}]{polar_Blumenson_1960}%
  \BibitemOpen
  \bibfield  {author} {\bibinfo {author} {\bibfnamefont {L.~E.}\ \bibnamefont
  {Blumenson}},\ }\bibfield  {title} {\bibinfo {title} {A derivation of
  n-dimensional spherical coordinates},\ }\href
  {http://www.jstor.org/stable/2308932} {\bibfield  {journal} {\bibinfo
  {journal} {The American Mathematical Monthly}\ }\textbf {\bibinfo {volume}
  {67}},\ \bibinfo {pages} {63} (\bibinfo {year} {1960})}\BibitemShut {NoStop}%
\bibitem [{Note6()}]{Note6}%
  \BibitemOpen
  \bibinfo {note} {Supplementary Material Sec XIII}\BibitemShut {NoStop}%
\bibitem [{Note7()}]{Note7}%
  \BibitemOpen
  \bibinfo {note} {We shall use the notation $\protect \mathcal {O}^{(k)}$ to
  denote the operator $\protect \mathcal {O}$ meant for qubits realized in the
  $k$-th 2-dimensional subspace. The operators for $d$-dimensional qudits are
  represented as $\protect \mathcal {O}^{[k]}$.}\BibitemShut {Stop}%
\bibitem [{\citenamefont
  {Skagerstam}(2018)}]{quantumClassicalEquiv_Skagerstam_2018}%
  \BibitemOpen
  \bibfield  {author} {\bibinfo {author} {\bibfnamefont {B.-S.~K.}\
  \bibnamefont {Skagerstam}},\ }\bibfield  {title} {\bibinfo {title} {On the
  three-slit experiment and quantum mechanics},\ }\href
  {https://doi.org/10.1088/2399-6528/aaf683} {\bibfield  {journal} {\bibinfo
  {journal} {Journal of Physics Communications}\ }\textbf {\bibinfo {volume}
  {2}},\ \bibinfo {pages} {125014} (\bibinfo {year} {2018})}\BibitemShut
  {NoStop}%
\bibitem [{\citenamefont {Sudarshan}(1963)}]{OET_Sudarshan_1963}%
  \BibitemOpen
  \bibfield  {author} {\bibinfo {author} {\bibfnamefont {E.~C.~G.}\
  \bibnamefont {Sudarshan}},\ }\bibfield  {title} {\bibinfo {title}
  {Equivalence of semiclassical and quantum mechanical descriptions of
  statistical light beams},\ }\href
  {https://doi.org/10.1103/PhysRevLett.10.277} {\bibfield  {journal} {\bibinfo
  {journal} {Phys. Rev. Lett.}\ }\textbf {\bibinfo {volume} {10}},\ \bibinfo
  {pages} {277} (\bibinfo {year} {1963})}\BibitemShut {NoStop}%
\bibitem [{\citenamefont {Ghosh}\ \emph {et~al.}(2018)\citenamefont {Ghosh},
  \citenamefont {Jennewein}, \citenamefont {Kolenderski},\ and\ \citenamefont
  {Sinha}}]{Spatial_Ghosh_2018}%
  \BibitemOpen
  \bibfield  {author} {\bibinfo {author} {\bibfnamefont {D.}~\bibnamefont
  {Ghosh}}, \bibinfo {author} {\bibfnamefont {T.}~\bibnamefont {Jennewein}},
  \bibinfo {author} {\bibfnamefont {P.}~\bibnamefont {Kolenderski}},\ and\
  \bibinfo {author} {\bibfnamefont {U.}~\bibnamefont {Sinha}},\ }\bibfield
  {title} {\bibinfo {title} {Spatially correlated photonic qutrit pairs using
  pump beam modulation technique},\ }\href
  {https://doi.org/10.1364/OSAC.1.000996} {\bibfield  {journal} {\bibinfo
  {journal} {OSA Continuum}\ }\textbf {\bibinfo {volume} {1}},\ \bibinfo
  {pages} {996} (\bibinfo {year} {2018})}\BibitemShut {NoStop}%
\bibitem [{\citenamefont {Walborn}\ \emph {et~al.}(2002)\citenamefont
  {Walborn}, \citenamefont {Terra~Cunha}, \citenamefont {P\'adua},\ and\
  \citenamefont {Monken}}]{customSlits_Monken_2002}%
  \BibitemOpen
  \bibfield  {author} {\bibinfo {author} {\bibfnamefont {S.~P.}\ \bibnamefont
  {Walborn}}, \bibinfo {author} {\bibfnamefont {M.~O.}\ \bibnamefont
  {Terra~Cunha}}, \bibinfo {author} {\bibfnamefont {S.}~\bibnamefont
  {P\'adua}},\ and\ \bibinfo {author} {\bibfnamefont {C.~H.}\ \bibnamefont
  {Monken}},\ }\bibfield  {title} {\bibinfo {title} {Double-slit quantum
  eraser},\ }\href {https://doi.org/10.1103/PhysRevA.65.033818} {\bibfield
  {journal} {\bibinfo  {journal} {Phys. Rev. A}\ }\textbf {\bibinfo {volume}
  {65}},\ \bibinfo {pages} {033818} (\bibinfo {year} {2002})}\BibitemShut
  {NoStop}%
\end{thebibliography}%
\nocite{customSlits_Monken_2002}

\clearpage




\section*{Supplementary material}

# Supplementary Material : Quantum State Interferography

Surya Narayan Sahoo,[1] Sanchari Chakraborti,[1] Arun K. Pati,[2] and Urbasi Sinha[1, *]

[1]*Light and Matter Physics, Raman Research Institute, Bengaluru 560080, India*
[2]*Quantum Information and Computation Group,
Harish-Chandra Research Institute, HBNI, Allahabad 211019, India*

## I. SCALING ADVANTAGE OF QUANTUM STATE INTERFEROGRAPHY

Quantum State Tomography (QST) is the standard method used for characterizing an unknown quantum state which involves a set of measurements to be performed on an ensemble of identically prepared copies of the quantum system and post processing of the collected data. For a $d$-dimensional system QST requires $d^2 - 1$ measurements when no assumptions are made about the system. Any prior knowledge of the state being pure reduces the required number of measurements in QST to $5d - 7$. Quantum State Interferography (QSI) is an interferometric method in which an unknown quantum state can be inferred from the visibilities, phase shifts and average intensities obtained from a number of interference patterns. Using this technique any unknown qubit ($d = 2$) state, be it mixed or pure, can be characterized from one interference pattern formed at the end of an interferometer having $\sigma_x$ operator in one arm and projection operator $\Pi_0$ in the other arm. Any pure state reconstruction in $d$-dimensions QSI requires $d - 1$ interferograms which can be obtained with two interferometers. Thus, for characterization of a state QSI, demands a number of measurements that scales linearly with the dimensionality of the system. In QST, on the other hand the required number of measurements scales quadratically with respect to the system size. This makes the process of QST more cumbersome as the dimensionality increases. The difference between the linear scaling in QSI and the quadratic scaling in QST can be understood better with the table below.

| $d$ | $n_1 = d^2 - 1$ | $n_2 = 5d - 7$ | $n_3 = d - 1$ | $\Delta n = n_1 - n_3$ | $\Delta n_p = n_2 - n_3$ |
|---|---|---|---|---|---|
| Dimension of system | QST without assumption | QST assuming pure state | QSI (pure state assumption for $d \geq 3$) | Difference in No. of required Measurements | Difference (with pure state assumption) |
| 2 | 3 | 3 | 1 | 2 | 2 |
| 3 | 8 | 8 | 2 | 6 | 6 |
| 4 | 15 | 13 | 3 | 12 | 10 |
| 5 | 24 | 18 | 4 | 20 | 14 |
| 10 | 99 | 43 | 9 | 90 | 34 |
| 50 | 2499 | 243 | 49 | 2450 | 194 |
| 100 | 9999 | 493 | 99 | 9900 | 394 |

TABLE I. Comparison of number of measurements (or experimental settings) required in QST vs. QSI

From the table, it is clear that as the dimensionality increases more and more, the difference between the required number of measurements in QST and QSI becomes more and more prominent. Even for characterizing qubit, QST requires three measurements to be performed whereas QSI suffices with only one measure-

* usinha@rri.res.in

ment. In case of characterizing qudits of large $d$ systems (with or without assumption of the state being pure) QSI sufficiently reduces the required number of measurements which can be seen from the numbers $\Delta n$ and $\Delta n_p$ in the table. The values $\Delta n_p$ for different $d$ makes it clear that even though the required number of measurements for QST with assumption of the state to be pure scales linearly with $d$, the number is still much higher compared to the number required in QSI when the system (or $d$) is sufficiently large. Thus, the process of Quantum State Interferography becomes more useful and advantageous for characterizing higher dimensional systems than the process of Quantum State Tomography from the view point of scaling gain.

Another advantage of QSI over QST can be described in terms of the need to change the experimental condition/setting in QST but not in QSI. A higher $d$-dimensional state characterization using QST requires measurements of $\mathcal{O}(d^2)$ to be performed which in turn requires the experimental condition to be changed for that many times. But the characterization of a $d$-dimensional state using QSI requires measurement of $d-1$ interferograms. These can be obtained using as low as 2 interferometers and without the need of any change in condition once and thus serve as single-shot method for the reconstruction of pure qudit states.

## II. DERIVATION OF INTENSITY AT THE OUTPUT PORT OF INTERFEROMETER

Any pure state for one-qubit can be written in terms of Bloch sphere coordinates as
$|\psi\rangle = \{\cos(\theta/2), \sin(\theta/2)\exp(i\phi)\}$. By knowing $\theta$ and $\phi$ from an experiment, we can infer any unknown pure state. If we compute the expectation value of spin ladder operator $\sigma_- = \frac{1}{2}(\sigma_x - i\sigma_y)$ in the state $|\psi\rangle$, we have the polar coordinate $\theta$ appearing only in the magnitude of the complex expectation value $\langle\sigma_-\rangle = (1/2)\sin(\theta)\exp(-i\phi)$ and the azimuthal coordinate $\phi$ appearing only as a phase in the Argand plane. The expectation value of the non-Hermitian spin ladder operator $\sigma_-$ cannot be obtained from statistics of measurement outcomes but can be inferred from the weak value [1] of $R$, where $R = \sqrt{\sigma_-^\dagger \sigma_-}$, in the pre-selected state $|\psi\rangle$ and the post-selected state $|\phi\rangle = U^\dagger |\psi\rangle$, where $U$ is a unitary matrix satisfying $\sigma_- = UR$. [2].

If we consider a mixed state for one-qubit (say a spin 1/2 particle), a complete description of its density matrix $\rho_s$ requires knowledge of three parameters $\theta, \phi$ and $\mu$.

$$\rho_s = \begin{pmatrix} \cos^2\left(\frac{\theta}{2}\right) & \frac{1}{2}\mu e^{-i\phi}\sin(\theta) \\ \frac{1}{2}\mu e^{i\phi}\sin(\theta) & \sin^2\left(\frac{\theta}{2}\right) \end{pmatrix} \quad (1)$$

Here, $\mu$ governs the purity of the density matrix $\rho_s$.

$$\text{Tr}(\rho_s^2) = \frac{1}{4}\left(3 + \mu^2 - \left(\mu^2 - 1\right)\cos(2\theta)\right) \quad (2)$$

The three (real) parameters for an unknown one-qubit mixed state can not be obtained from the weak value (a single complex number). However, it has been shown that the expectation value of $\sigma_-$ in a pure state $|\psi\rangle$ or the weak value of $R$ in the preselected state $|\psi\rangle$ and postselected state $|\phi\rangle = U^\dagger |\psi\rangle$ can be inferred from visibility and phase shift obtained in a Mach-Zehnder interferometer [3]. In this article, we show that any one-qubit mixed state $\rho_s$ can be inferred from the phase-averaged intensity, visibility and phase shift of the interference pattern obtained in the Mach-Zehnder interferometer, where the optical component corresponding to $R$ is the transmitting port of a polarizing beam splitter (PBS) or polarizer with its transmission axis oriented along horizontal and it is placed in one arm of the interferometer and the optical component corresponding to $U$ (the half-wave plate (HWP) with its fast axis oriented at an angle $\pi/4$ from horizontal i.e as $\sigma_x$) is placed in the other arm.

Let us consider the experimental set-up, as shown in Fig. 1. In the polarization subspace, the Jones representation for a Polarizer with transmission axis oriented along horizontal (or for a Polarizing Beam Splitter) becomes a non-unitary matrix. Hence, the overall evolution operator that describes the Mach-Zehnder interferometer, along with components shown in Fig. 1, is not trace-preserving. To have a unitary description, we have to include all the states in the path degree of freedom as well and describe all the operators in a higher dimensional Hilbert space.

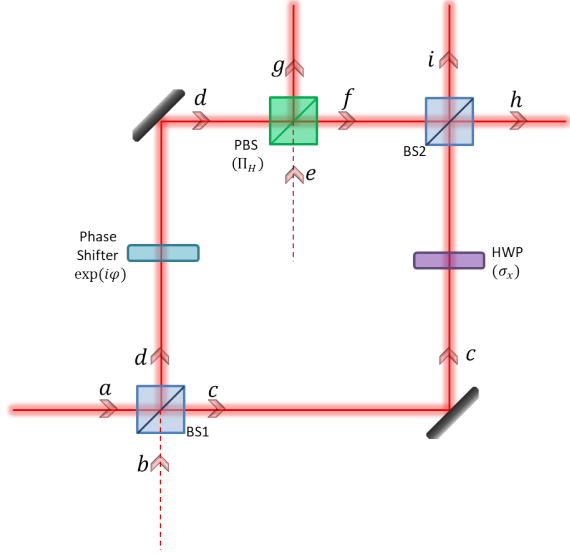

FIG. 1. Mach Zehnder Set-up for Polarization State Interferography

There are 3 input paths to the Mach-Zehnder interferometer $\{|a\rangle, |b\rangle, |e\rangle\}$. The beam is made incident only in port $|a\rangle$ and hence the density matrix for path degree of freedom in the basis $\{|a\rangle, |b\rangle, |e\rangle\}$ is given as follows:

$$\rho_p = \begin{pmatrix} 1 & 0 & 0 \\ 0 & 0 & 0 \\ 0 & 0 & 0 \end{pmatrix} \quad (3)$$

The incident beam may then be represented as an outer product of the path and polarization (spin) density matrices.

$$\rho_i = \rho_p \otimes \rho_s = \begin{pmatrix} \cos^2\left(\frac{\theta}{2}\right) & \frac{1}{2}\mu e^{-i\phi}\sin(\theta) & 0 & 0 & 0 & 0 \\ \frac{1}{2}\mu e^{i\phi}\sin(\theta) & \sin^2\left(\frac{\theta}{2}\right) & 0 & 0 & 0 & 0 \\ 0 & 0 & 0 & 0 & 0 & 0 \\ 0 & 0 & 0 & 0 & 0 & 0 \\ 0 & 0 & 0 & 0 & 0 & 0 \\ 0 & 0 & 0 & 0 & 0 & 0 \end{pmatrix} \quad (4)$$

Note that the above representation is in the basis $\{|a\rangle, |b\rangle, |e\rangle\} \otimes \{|H\rangle, |V\rangle\}$.

The first beam-splitter transforms the input ports $\{|a\rangle, |b\rangle\}$ to the ports $\{|c\rangle, |d\rangle\}$ and leaves the port $\{|e\rangle\}$ unaffected. The second beam-splitter transforms $\{|c\rangle, |f\rangle\}$ to the output ports $\{|i\rangle, |h\rangle\}$ and leaves the port $|g\rangle$ unaffected. The matrix representations of the beam-splitter operations in the path subspace are as follows:

$$b_1 = \frac{1}{\sqrt{2}}\begin{pmatrix} 1 & i & 0 \\ i & 1 & 0 \\ 0 & 0 & \sqrt{2} \end{pmatrix} \quad b_2 = \frac{1}{\sqrt{2}}\begin{pmatrix} 1 & i & 0 \\ i & 1 & 0 \\ 0 & 0 & \sqrt{2} \end{pmatrix} \quad (5)$$

The complete beam splitter operators in the path and polarization degree of freedom are given by $B_1 = b_1 \otimes \mathbb{1}$ and $B_2 = b_2 \otimes \mathbb{1}$ respectively.

The phase-shifter acts only on path $|d\rangle$ and hence we write the operator as $U(\Phi) = (\Pi_d \exp(i\varphi) + (\Pi_c + \Pi_e)) \otimes \mathbb{1}$.

The action of the half-wave plate can be viewed as the $\sigma_x$ operation in arm $|c\rangle$ and identity on $\{|d\rangle, |e\rangle\}$.

$$H = \Pi_c \otimes \sigma_x + (\mathbb{1}^{(3)} - \Pi_c) \otimes \mathbb{1}^{(2)} = \begin{pmatrix} 0 & 1 & 0 & 0 & 0 & 0 \\ 1 & 0 & 0 & 0 & 0 & 0 \\ 0 & 0 & 1 & 0 & 0 & 0 \\ 0 & 0 & 0 & 1 & 0 & 0 \\ 0 & 0 & 0 & 0 & 1 & 0 \\ 0 & 0 & 0 & 0 & 0 & 1 \end{pmatrix} \quad (6)$$

The matrix for the polarizing beam splitter is the transformation from the basis $\{|d\rangle, |e\rangle\}$ to $\{|f\rangle, |g\rangle\}$ and the port $|c\rangle$ remains identical.

$$P = \begin{pmatrix} 1 & 0 & 0 & 0 & 0 & 0 \\ 0 & 1 & 0 & 0 & 0 & 0 \\ 0 & 0 & 1 & 0 & 0 & 0 \\ 0 & 0 & 0 & 0 & 0 & 1 \\ 0 & 0 & 0 & 0 & 1 & 0 \\ 0 & 0 & 0 & 1 & 0 & 0 \end{pmatrix} \quad (7)$$

Now, we can compute the final density matrix after the Mach-Zehnder interferometer.

$$\rho_{\{|i\rangle,|h\rangle,|g\rangle\}\otimes\{|H\rangle,|V\rangle\}}^{B_1 \to \Phi \to H \to P \to B_2 \to}$$
$$= B_2 \cdot (P \cdot (H \cdot (\Phi \cdot (B_1 \cdot \rho_i \cdot B_1^\dagger) \cdot \Phi^\dagger) \cdot H^\dagger) \cdot P^\dagger) \cdot B_2^\dagger \quad (8)$$

Till now, all the evolution has been unitary and the above density matrix has unit trace.

We now place the detector/beam-profiler only in the port $\{|h\rangle\}$. We represent the projector as $\Pi_h$.

$$\Pi_h = \begin{pmatrix} 0 & 0 & 0 & 0 & 0 & 0 \\ 0 & 0 & 0 & 0 & 0 & 0 \\ 0 & 0 & 1 & 0 & 0 & 0 \\ 0 & 0 & 0 & 1 & 0 & 0 \\ 0 & 0 & 0 & 0 & 0 & 0 \\ 0 & 0 & 0 & 0 & 0 & 0 \end{pmatrix} \quad (9)$$

The resultant component of the density matrix in the port $|h\rangle$ that is going to be detected is given by the following reduced density matrix

$$\rho_d = \begin{pmatrix} \frac{1}{4}(1 + \mu \sin(\theta) \cos(\phi - \varphi)) & \frac{1}{8}\left(\mu e^{i\phi} \sin(\theta) + e^{i\varphi}(\cos(\theta) + 1)\right) \\ \frac{1}{8}\left(\mu e^{-i\phi} \sin(\theta) + e^{-i\varphi}(\cos(\theta) + 1)\right) & \frac{1}{8}(\cos(\theta) + 1) \end{pmatrix} \quad (10)$$

The total intensity at the detector is obtained by taking the trace of the above density matrix in the polarization subspace, i.e.,

$$I_d = \text{Tr}(\rho_d) = \frac{1}{8}(3 + \cos(\theta) + 2\mu \sin(\theta) \cos(\phi - \varphi)). \quad (11)$$

The phase shift ($\Phi$) of the interference pattern is obtained at the value of $\varphi$ that maximizes $I_d$. Since $0 \leq \theta \leq \pi$ in a Bloch Sphere, sine in the last term is always positive and hence the phase shift is given by $\phi$. The phase averaged power $\bar{I}$ is given by $\frac{1}{8}(3 + \cos(\theta))$.

Thus, the $\theta, \phi$ are determined from the average power and the phase shift in an interferometer port.

Once $\theta$ is known, $\mu$ can be obtained from visibility $V$ which is given by $\frac{2\mu \sin(\theta)}{3+\cos(\theta)}$.

## III. PHASE SHIFT, VISIBILITY AND AVERAGE INTENSITY AS FUNCTION OF BLOCH SPHERE PARAMETERS

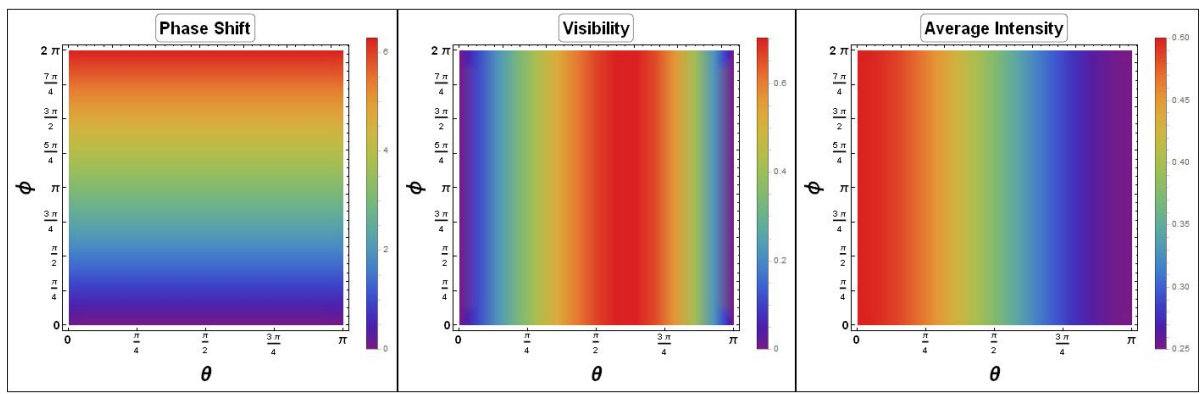

FIG. 2. Phase Shift, Visibility and Average Intensity as function of Bloch Sphere Parameters: The phase shift obtained is $\theta$ independent and is a unique function of $\phi$. Visibility however, although independent of $\phi$ is a many(two) to one function of $\theta$. Therefore, we consider the phase averaged intensity which is a one to one map of $\theta$. Visibility, nevertheless, helps in distinguishing $\mu$ - the parameter governing purity of the state.

## IV. COMPARISON OF VARIOUS INTERFEROMETRIC METHODS

In the experiment we want to determine the phase shift, visibility and the average intensity from the interference pattern obtained at one of the output ports of an interferometer in order to reconstruct the density matrix associated with the polarization state incident on the interferometer. This input state is generated by the action of a HWP (at angle $\alpha$) followed by a QWP (at angle $\beta$) on the vertical polarization state emitted from the source. In order to obtain the phase shift of the interference pattern as one changes $\alpha$ and $\beta$ the interferometer needs to be phase-stabilized.

One of the disadvantages of using the Mach Zehnder interferometer is that we need to stabilize the path difference against external vibrations to obtain any consistent phase information. Hence, to infer phase shift, we prefer interferometers that are not prone to vibrations, e.g., double slit interferometer and the Sagnac interferometer. We can simply use the equivalent double slit set up, with one slit attached with a polarizer (R) and the other slit filled with the half wave plate (U). The interference pattern here would be insensitive to external noise because the inter-slit distance is robust to noise and thus acts as an ideal device to give us the interference pattern from which both visibility and phase shift can be accurately obtained.

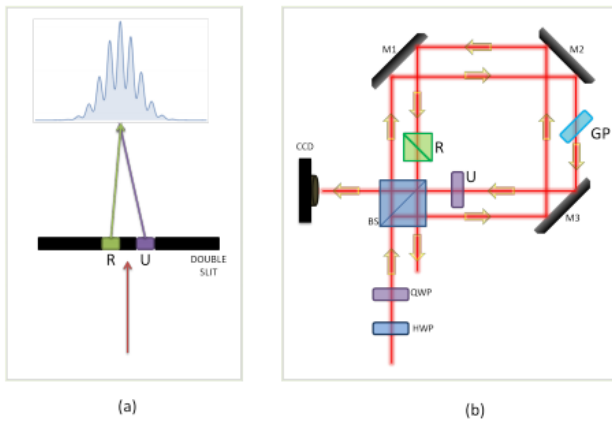

FIG. 3. Double Slit and Sagnac interferometer as alternative to Mach Zehnder interferometer

To go for a miniaturized setup it is best to custom design [4] a slit version wherever possible. But, the polarizer and the half wave-plate for the double-slit needs to be specifically manufactured and placed appropriately on the slits. Any difference in shape of slits and that of polarizer/waveplate would lead to additional diffraction effects. However, with a displaced Sagnac geometry, we can achieve the phase stability against external low frequency vibrations, with components readily available in an optics lab.

## V. DATA ANALYSIS & CHOICE OF STATISTICS

For a fixed angle $\alpha$ of the HWP, we rotate the QWP and obtain 5 images for a given $\beta$. We first orient the image to have fringes along the horizontal (along $x_f$). For each image, we take 100 horizontal slices about the vertical centroid and fit each slice with the model which is a Gaussian weighted cosine function given below.

$$B_f + A_f \exp\left(-c_f(x_f - m_f)^2\right)(1 + v_f \cos(k_f x_f + \phi_f)). \tag{12}$$

Here, $B_f$ is the background noise and $A_f$ is the amplitude of the Gaussian envelope centred at $m_f$ with standard deviation as $\sqrt{1/2c_f}$. The fringe width is given by $2\pi/k_f$. The visibility of the fringe and phase shift are determined from $v_f$ and $\phi_f$ respectively.

We weigh the data with the vertical Gaussian profile and then take mean and standard deviation of each of the parameters over the 100 slices. If the model does not fit any of the slices, i.e., the adjusted $R^2$ of the fit was less than 0.99, we give it zero weight. For $\phi_f$, we use circular mean and circular standard deviation (See definition in VI) . The amplitude $A_f$ is corrected against the vertical Gaussian weight. Finally, we take the average (over 5 images) of the the averages (over 100 slices) to represent the mean experimental value. The error bars are represented by the maximum of standard deviation (over 5 images) of means (over 100 slices) and the RMS value of the standard deviations (over 100 slices). We then repeat the same for various HWP angles $\alpha$. Finally, we take a dataset with QWP absent in order to find the the zero reference of the phase shift.

The interference pattern in a Sagnac interferometer is not affected by the low frequency vibrations and hence repetitions of image acquisition over time are reproducible. Thus taking five images (exposure of 13 ms)

in intervals of about 500 ms for a given prepared state give us consistent results. However, the envelope of the transverse profile of the beam is not perfectly Gaussian because of some (avoidable) dust on the optical components and on the imaging sensor. Therefore, instead of taking just one profile from the image, we take 100 horizontal slices about the vertical centroid. The slices which are far off the vertical centroid are affected the most due to lack of perfect overlap between the two beams coming from two different paths of the interferometer. Therefore, 100 slices, which are within FWHM of the vertical Gaussian envelope, is optimal to have enough sampling of the beam to eliminate irregularities. The standard deviation of the fit parameters is obtained because the variation of these 100 slices is larger than the changes in mean of the fit parameters across 5 images and hence 5 images give us enough statistics.

## VI. CIRCULAR MEAN AND STANDARD DEVIATION

Consider an array $A_\phi$ containing the list of circular variable $\phi$. Then the circular mean for a list of $N$ values is given by

$$\mu_c^\phi = \arctan\left(\frac{\sum_i^N \cos\left(A_\phi^i\right)}{N}, \frac{\sum_i^N \sin\left(A_\phi^i\right)}{N}\right). \quad (13)$$

Here, $\arctan(x, y)$ gives the arc tangent of $y/x$ after taking into account the quadrant to which the pair belongs to.

We use the square root of the circular variance as the standard deviation which is computed as follows:

$$\sigma_c^\phi = \sqrt{1 - \frac{\sqrt{\left(\sum_i^N \cos\left(A_\phi^i\right)\right)^2 + \left(\sum_i^N \sin\left(A_\phi^i\right)\right)^2}}{N}} \quad (14)$$

## VII. RESULTS

From the interference pattern obtained in the non-collinear displaced Sagnac interferometer, we obtain the phase shift, the visibility and the average intensity for different polarization states prepared by different $(\alpha, \beta)$ combinations as described in the previous section. Then compute the fidelity of the reconstructed states compared to the prepared ones.

In the experiment, the state is prepared by using a HWP at angle $\alpha$ followed by QWP at angle $\beta$ on the incident polarization state $V$.

Thus, the prepared state $|\psi(\alpha, \beta)\rangle$ can be expressed as follows:

$$|\psi(\alpha, \beta)\rangle = \text{QWP}(\beta) \cdot \text{HWP}(\alpha) \cdot |V\rangle \quad (15)$$

$$= \begin{pmatrix} \cos^2(\beta) + \imath \sin^2(\beta) & (1-\imath)\sin(\beta)\cos(\beta) \\ (1-\imath)\sin(\beta)\cos(\beta) & \sin^2(\beta) + \imath \cos^2(\beta) \end{pmatrix} \cdot \begin{pmatrix} \cos(2\alpha) & \sin(2\alpha) \\ \sin(2\alpha) & -\cos(2\alpha) \end{pmatrix} \cdot \begin{pmatrix} 0 \\ 1 \end{pmatrix} \quad (16)$$

$$= \frac{1}{2} \begin{pmatrix} (1+\imath)(\sin(2\alpha) - \imath \sin(2\alpha - 2\beta)) \\ (1-\imath)(\imath \cos(2\alpha) + \cos(2\alpha - 2\beta)) \end{pmatrix} = \begin{pmatrix} \psi_H \\ \psi_V \end{pmatrix}; \text{say} \quad (17)$$

The relation between $\theta, \phi$ used in the Eqn. 1 with $\alpha, \beta$ is shown below:

$$\theta = 2 * \arccos\left(\psi_H e^{(-\imath \arg(\psi_H))}\right) \quad (18)$$

$$\phi = -\imath \ln\left(\frac{\psi_V e^{(-\imath \arg(\psi_H))}}{\sin(\theta/2)}\right) \quad (19)$$

The variations of the phase shift, the averaged intensity and the visibility of the interference pattern with the HWP angle ($\alpha$) and the QWP angle ($\beta$) have been shown in Fig. 4, Fig. 5, Fig. 6 respectively . The solid lines in the plots represent the theoretical prediction while the dots and bars represent experimentally obtained mean and statistical error respectively. The black curve is for the experiment where only HWP was rotated in absence of the QWP.

The HWP does not introduce any phase shift, i.e., it transforms an incident linear polarization to another linear polarization. However, as the two interfering beams have arbitrary but constant path difference in the displaced Sagnac geometry, the phase shift obtained is not an exact zero but some constant, which is then taken as the reference. The phase shift obtained from the interferogram has more error when $\theta$ is closer to 0 or $\pi$, since the Bloch vector is closer to the poles where $\phi$ is undefined. This is manifested in noticeable deviations of the experimental graphs from the theory for HWP angles 0 and 45.

All the experimentally obtained averaged intensity are normalized (with norm = 0.5) with respect to the corresponding maximum of the average intensity obtained as a function of HWP in the absence of QWP. This normalization step can be avoided if the detector is calibrated against known input intensity. The average intensity only depends on the polarization state and not on the spatial overlap of the beams, stability of the interferometer, wavefront or beam shape.

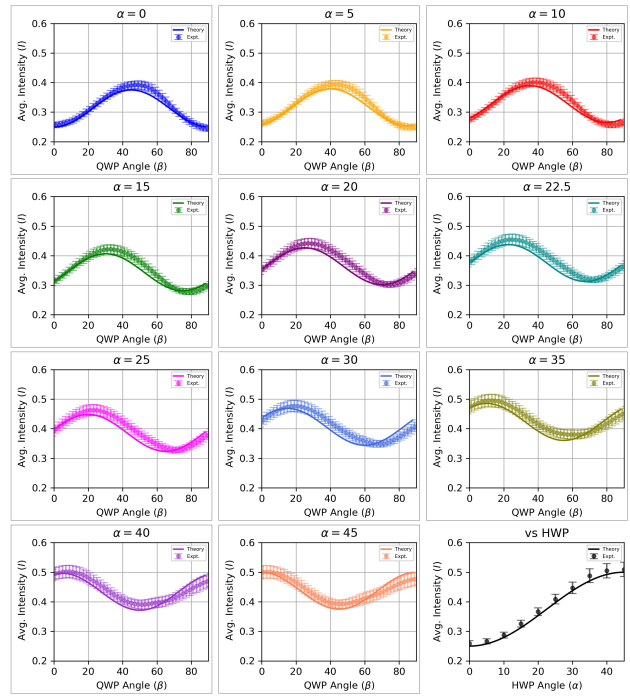

FIG. 5. Average Intensity

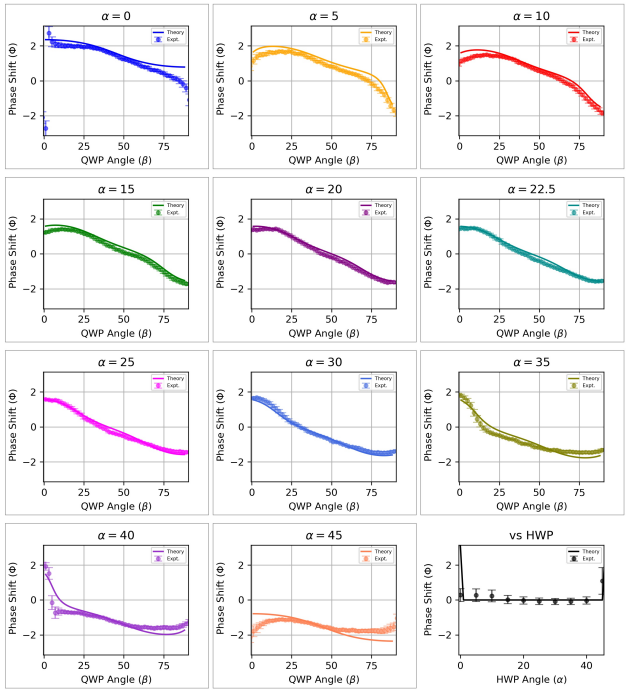

FIG. 4. Phase Shift:

The particular HWP and QWP used cause angular deviation of the beam as they are rotated that results in change in the overlap of the two beams at the detector. The beam splitter also has about 3% polarization-dependent transmission and reflection probability. These effects, along with intensity averaging over the area of the sensor contribute to the experimentally obtained visibility being systematically lower than the theory at certain angles of the wave plate.

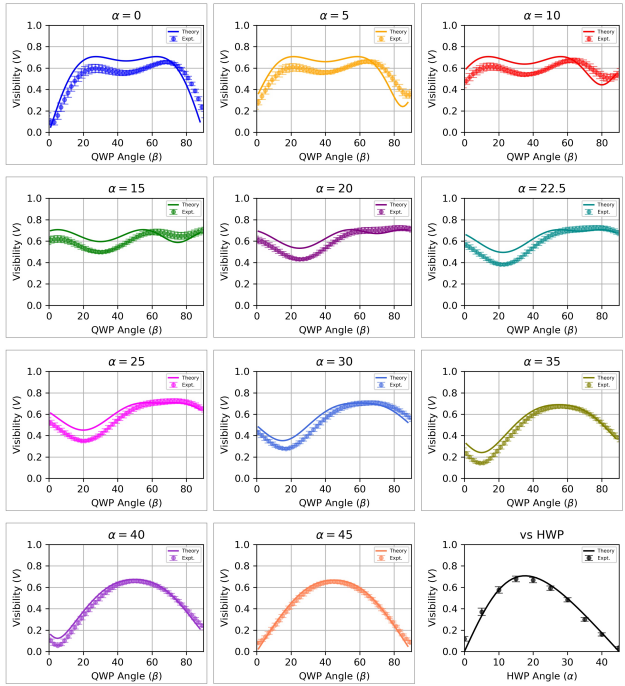

FIG. 6. Visibility

## VIII. PHYSICALITY OF THE RECONSTRUCTED DENSITY MATRIX

The density matrix in two-dimensions in terms of Bloch sphere coordinates $\theta$ and $\phi$ is represented as,

$$\rho = \begin{pmatrix} \cos^2\left(\frac{\theta}{2}\right) & \frac{1}{2}\mu e^{-i\phi}\sin(\theta) \\ \frac{1}{2}\mu e^{i\phi}\sin(\theta) & \sin^2\left(\frac{\theta}{2}\right) \end{pmatrix}. \qquad (20)$$

For all values of $\theta$ and $\phi$, the density matrix is physical by virtue of construction for $0 \leq \mu \leq 1$. After we find out $\phi$ and $\theta$, we have $\mu$ from visibility as

$$\mu = V\left(\frac{3 + \cos(\theta)}{2\sin(\theta)}\right) \qquad (21)$$

When $\theta$ is close to 0 or $\pi$, the denominator tends to 0. At this point experimentally obtained visibility has to be low enough so that $\mu \leq 1$. Due to experimental imperfections and noise, experimentally obtained visibility sometimes can be slightly higher or lower than what one would expect with ideal experimental components. In this sense, by simple use of the formula in Eqn. (21), one could obtain $\mu > 1$. However, note that this occurs when the state is close to $|H\rangle$ or $|V\rangle$, i.e. when $\theta = 0$ or $\pi$ and hence the off-diagonal terms in $\rho$ which contains $\sin(\theta)$ tends to 0. Thus, if we clip the value of $\mu$ between 0 and 1, the error in $\rho$ is minimal. Next, we shall quantify this error. But before that, in an example, we show that in our experiment $\mu$ is almost always found to be within 1, with only few exceptions. For instance, when the state was prepared with $HWP$ angle $\alpha = 0$, we plot $\mu$ obtained as a function of prepared state parameterized by the QWP angle $\beta$.

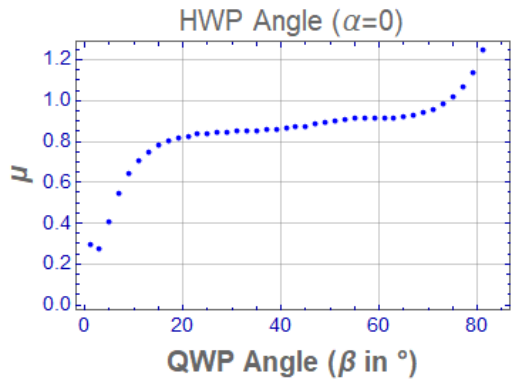

FIG. 7. $\mu$ obtained as a function of prepared state parameterized by the QWP angle $\beta$ for the HWP angle $\alpha = 0$. Except few points in the region where $\beta$ is close to 90°, we find $\mu < 1$ and hence the density matrix is physical. Note that the value is around 0.8 and 0.9 for most values of $\beta$ i.e. $15° < \beta < 75°$. It drops sharply towards 0 as $\beta \to 0$ (See Sec. (VII) for relation between $(\theta, \phi)$ and $(\alpha, \beta)$ ). Again, the state approaches $|V\rangle$ (or $|H\rangle$) and hence the low value of $\mu$ do not contribute to density matrix being different.

For the few points, where $\mu > 1$, we have coerced the value to 1 by taking $\mu \equiv \min(\mu, 1)$. Again emphasis is to be made that this does not affect $\rho$ much because the off-diagonal terms are anyways small because of $\theta$ being closer to 0 or $\pi$. Anyways, we outline a generic procedure in which $\mu$ can be followed to systematically obtain $\mu$ without the need of clipping the value between 0 and 1.

In the Eqn. (9), of the main text, the function which used to fit the experimentally obtained interference pattern is provided. Instead of finding the best fit for the parameters for $v_f$ and $A_f$ along with other parame-

ters, we can directly substitute $v_f = \frac{2\mu \sin(\theta)}{3+\cos(\theta)}$ using Eqn. (8) and $A_f = \mathcal{N}\frac{1}{8}(3+\cos(\theta))$. Here, $\mathcal{N}$ is normalization/scaling factor associated with intensity measurement and is determined from experimental setup. In this way, now we can directly fit for $\theta$ and $\mu$ instead of $A_f$ and $v_f$. The fitting algorithm is essentially a constrained optimization with the bounds for $\theta$ being $[0, \pi)$ and that for $\mu$ being within $[0, 1]$. Thus, the requirement that $0 \leq \mu \leq 1$ can be directly imposed in the fitting algorithm rather than imposing the constraints on $v_f$.

It is also worth noting that the reconstructed density matrix from Quantum State Tomography also requires additional post-processing so that the density matrix becomes physical [5, 6].

In experiment, the state is not always pure. We parameterize the mixedness with $\mu$ and show how much the fidelity of $\rho_{pure}$ (the density matrix for the pure state i.e. $\rho_{pure} = \rho\big|_{\mu=1}$) with $\rho$ changes as a function of $\mu$.

The fidelity is obtained as

$$\mathcal{F} = \text{Tr}\left(\sqrt{\sqrt{\rho_{pure}}\rho\sqrt{\rho_{pure}}}\right)^2 \quad (22)$$

$$= \frac{1}{4}(3 + \mu - (\mu-1)\cos(2\theta)) \quad (23)$$

Note, that the fidelity is independent of $\phi$ and thus mixedness of state does not affect the determination of the phase $\phi$. We plot the fidelity as function of $\mu$ and $\theta$ in Fig. 8. We observe that fidelity is close to 1 irrespective of the value of $\mu$ when $\theta$ is close to 0 or $\pi$. Only when $\theta$ is around $\pi/2$ the fidelity drops when mixedness increases very much i.e., $\mu << 1$. For a small amount of mixedness due to errors in preparation of state, we have $\mathcal{F} > 0.9$ for $\mu > 0.8$. However, the experimentally obtained $\mu$ has also less errors when $\theta$ is close to $\pi/2$ as can be inferred from the Fig. 7 and therefore the fidelity obtained in the experiment is high.

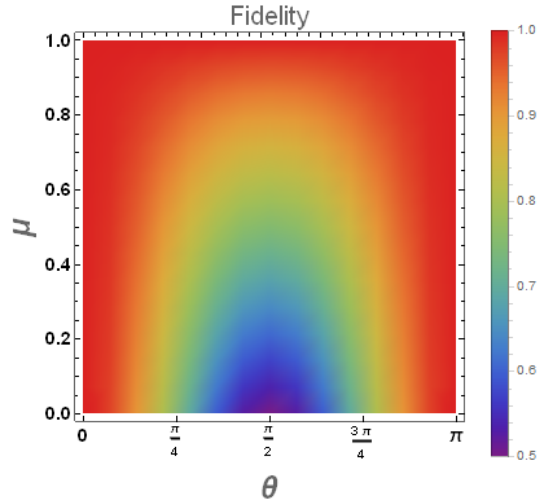

FIG. 8. Density plot of Fidelity between ideal pure state and mixed state parameterized by $\mu$. Note that the colors scale from 0.5(blue) to 1(red)
.

## IX. FIDELITY AND PURITY

We report median analysis as the errors introduced are state dependent. For instance, there is more uncertainty in determining phase shift whenever the visibility is low. The best case suggests that if systematic and random errors can be minimized (eg: by using cage mount assembly for better stability and miniaturizing the setup to avoid effects due to pointing fluctuations/ angular deviation caused due to waveplates), the method can give us state estimation with fidelity better than 0.98 (See Table II below). The median of the average mixed state fidelity over all the states is lower than the corresponding median for the reconstructed pure states. This is because of the errors in determining $\mu$ from the visibility which is affected by pixel averaging, overlap of the two beams and change in ellipticity of polarization at each reflection. Since we do not need visibility to reconstruct pure states, these errors only affect reconstruction of mixed states. However, errors such as change in polarization due to reflection can be avoided in the miniaturized slit based QSI setup discussed in Sec. IV .

|  | Average | Best Case | Worst Case |
|---|---|---|---|
| Purity ($\text{Tr}(\rho^2)$) | $0.92(5)^{+0.03(7)}_{-0.03(7)}$ | $0.98(0)^{+0.01(9)}_{-0.02(7)}$ | $0.85(7)^{+0.06(5)}_{-0.03(1)}$ |
| Mixed state Fidelity $\mathcal{F}_m$ | $0.94(1)^{+0.02(5)}_{-0.01(3)}$ | $0.98(1)^{+0.00(9)}_{-0.01(2)}$ | $0.90(2)^{+0.03(2)}_{-0.01(4)}$ |
| Pure state Fidelity $\mathcal{F}_p$ | $0.98(3)^{+0.00(4)}_{-0.00(6)}$ | $0.99(5)^{+0.00(2)}_{-0.00(5)}$ | $0.97(0)^{+0.00(6)}_{-0.00(6)}$ |

TABLE II. Purity and Fidelity: In this table, we report the median over all the prepared states of average fidelity along with the upper and lower quartile deviations. The same is also presented for the best case and worst case defined by highest and lowest fidelity (or purity) obtained after error (obtained from statistics of the average intensity, phase shift and visibility) propagation of half standard deviation about the mean quantities.

## X. REPRESENTATION OF QUDITS

We aim to extend the idea implemented for qubits for the reconstruction of the unknown state incident on the interferometer to $d$-dimensional qudits.

Any ray in $d$-dimensional Hilbert space can be parameterized by $\{\theta_j, \phi_j\}; j = 1, 2, \ldots, d-1$. The first term of the vector can always be made a positive real number because we can ignore the global phase and hence can always be expressed as $\cos(\theta_1/2)$. The second term can be in general complex but can have the maximum magnitude as $\sin(\theta_1/2)$. Hence in the polar form [7] it can be written as $\sin(\theta_1/2)\cos(\theta_2/2)\exp(i\phi_1)$. Similarly, the $k$-th term can be written as $\prod_{j=1}^{k-1} \sin\left(\frac{\theta_j}{2}\right) \exp(i\phi_j) \cos\left(\frac{\theta_k}{2}\right)$. The final term's magnitude is determined by all the previous terms due to normalization condition and hence is given by $\prod_{j=1}^{d-1} \sin\left(\frac{\theta_j}{2}\right) \exp(i\phi_j)$. Hence, the polar representation of a state in $d$-dimensional Hilbert space will be,

$$|\psi\rangle^{(d)} = \begin{pmatrix} \cos\left(\frac{\theta_1}{2}\right) \\ \sin\left(\frac{\theta_1}{2}\right) e^{i\phi_1} \begin{pmatrix} \cos\left(\frac{\theta_2}{2}\right) \\ \sin\left(\frac{\theta_2}{2}\right) e^{i\phi_2} \begin{pmatrix} \ddots \\ \sin\left(\frac{\theta_k}{2}\right) e^{i\phi_k} \begin{pmatrix} \cos\left(\frac{\theta_k}{2}\right) \\ \ddots \begin{pmatrix} \cos\left(\frac{\theta_{d-1}}{2}\right) \\ \sin\left(\frac{\theta_{d-1}}{2}\right) e^{i\phi_{d-1}} \end{pmatrix} \end{pmatrix} \end{pmatrix} \end{pmatrix} \end{pmatrix} \quad (24)$$

We can omit the nested brackets and write the state as the following single column vector.

$$|\psi\rangle^{(d)} = \begin{pmatrix} \cos\left(\frac{\theta_1}{2}\right) \\ \sin\left(\frac{\theta_1}{2}\right) \exp(i\phi_1) \cos\left(\frac{\theta_2}{2}\right) \\ \vdots \\ \prod_{j=1}^{k-1} \sin\left(\frac{\theta_j}{2}\right) \exp(i\phi_j) \cos\left(\frac{\theta_k}{2}\right) \\ \vdots \\ \prod_{j=1}^{d-1} \sin\left(\frac{\theta_j}{2}\right) \exp(i\phi_j) \end{pmatrix} \quad (25)$$

We prove that the coordinates $\{\theta_j, \phi_j\}; j = 1, 2 \ldots d-1$ makes $|\psi\rangle^{(d)}$ span the entire vector space by the use of principle of mathematical induction. We have already verified that this representation spans the Hilbert space for $d = 2$ (Bloch sphere for qubits) and $d = 3$ (two sequential Bloch vectors for qutrit). We then assume that $|\psi\rangle^{(k)}$ spans the $k$-dimensional Hilbert space and aim to argue that, by implication, we can conclude that $|\psi\rangle^{(k+1)}$ spans all the rays in the $k+1$ dimensional Hilbert space.

We have

$$|\psi\rangle^{(k+1)} = \begin{pmatrix} \alpha_1 \\ \vdots \\ \alpha_k \\ \alpha_{k+1} \end{pmatrix}$$

$$= \begin{pmatrix} \cos\left(\frac{\theta_1}{2}\right) \\ \vdots \\ \prod_{j=1}^{k-1} \sin\left(\frac{\theta_j}{2}\right) e^{i\phi_j} \begin{pmatrix} \cos\left(\frac{\theta_k}{2}\right) \\ \sin\left(\frac{\theta_k}{2}\right) e^{i\phi_k} \end{pmatrix} \end{pmatrix} \quad (26)$$

Since $|\psi\rangle^{(k)}$ spans the $k$-dimensional Hilbert space, we have $\{\alpha_1, \alpha_2 \ldots \alpha_k\}$ as the set of arbitrary complex numbers up to a global phase and constrained to normalization. Thus, we now have to show that $\alpha_{k+1}$ can be an arbitrary complex number. Because $\theta_k \in [0, \pi]$ and $\phi_k \in (-\pi, \pi]$ makes $\prod_{j=1}^{k-1} \sin\left(\frac{\theta_j}{2}\right) e^{i\phi_j} \sin\left(\frac{\theta_k}{2}\right) e^{i\phi_k}$ an arbitrary complex number with magnitude bounded by $|\prod_{j=1}^{k} \sin\left(\frac{\theta_j}{2}\right) e^{i\phi_j}|$, $|\psi\rangle^{(k+1)}$ spans the entire $k+1$-dimensional Hilbert space. Note that $\alpha_k$ has to be scaled by $\cos\left(\frac{\theta_k}{2}\right)$ but still represents an arbitrary complex number.

## XI. EXPERIMENTAL PROTOCOL

The proposal is to use the $d-1$ Mach Zehnder (or equivalent) Interferometers on each of the two dimensional $\{|k\rangle, |k+1\rangle\}$ subspace of the $d$-dimensional state $|\psi\rangle^{(d)}$ in the sequence in which we represent the vectors.

The $k$-th 2-dimensional subspace is given by

$$|\psi\rangle_k^{(2;d)} =$$
$$\left(\prod_{j=1}^{k-1} \sin\left(\frac{\theta_j}{2}\right) \exp(i\phi_j)\right) \begin{pmatrix} \cos\left(\frac{\theta_k}{2}\right) \\ \sin\left(\frac{\theta_k}{2}\right) \exp(i\phi_k) \cos\left(\frac{\theta_{k+1}}{2}\right) \end{pmatrix} \quad (27)$$

Next, we compute the expectation value of the spin ladder operator in the two dimensional subspace.

$$\langle\psi|\sigma_\pm|\psi\rangle_k^{(2;d)}$$
$$= \prod_{j=1}^{k-1} \sin^2\left(\frac{\theta_j}{2}\right) \left(\frac{1}{2} e^{\pm i\phi_k} \cos\left(\frac{\theta_{k+1}}{2}\right) \sin(\theta_k)\right) . \quad (28)$$

Using this we directly obtain the relative phase $\phi_k$ in the two-dimensional subspace from the argument of the expectation value of the spin ladder operator. To determine $\theta_k$, we need to know $\cos\left(\frac{\theta_{k+1}}{2}\right)$ as well as $\prod_{j=1}^{k-1} \sin^2\left(\frac{\theta_j}{2}\right)$ and the method to obtain the same is discussed in the next section.

Now, we discuss a generic protocol for measuring the expectation values of the sequence of spin ladder operators. We employ the same scheme of polar decomposing the ladder operator in $\Pi_0^{(k)}$ and $\sigma_x^{(k)}$ operators in the two-dimensional $k$-th subspace. On one arm, we have the operator $\exp(i\varphi_k)\Pi_0^{(k)}$, i.e., the phase shifter for the $k$-th MZI and projector on the mode $|k\rangle$ and the other arm having the $\sigma_x^{(k)}$ the spin-flip operator in the subspace that swaps $|k\rangle$ with $|k+1\rangle$. We have to design $d-1$ such MZI setup for a $d$-dimensional qudit. In each such MZI, we are effectively obtaining the expectation value of the ladder operator for $k$ to $k+1$ mode.

First, we divide the beam into $d-1$ spatial modes. For qutrit, we can use a 50:50 Beam splitter $BS_0$ to do the same. For each beam, we construct the Mach-Zehnder interferometer which acts on a 2-dimensional subspace. Here, the beam splitters $BS_{11}$ and $BS_{12}$ forms the MZI for the subspace consisting of spin modes $\{|0\rangle, |1\rangle\}$. This is achieved by splitting the beam into three spin modes by the Spin Tritter $ST$ and blocking $|2\rangle$.

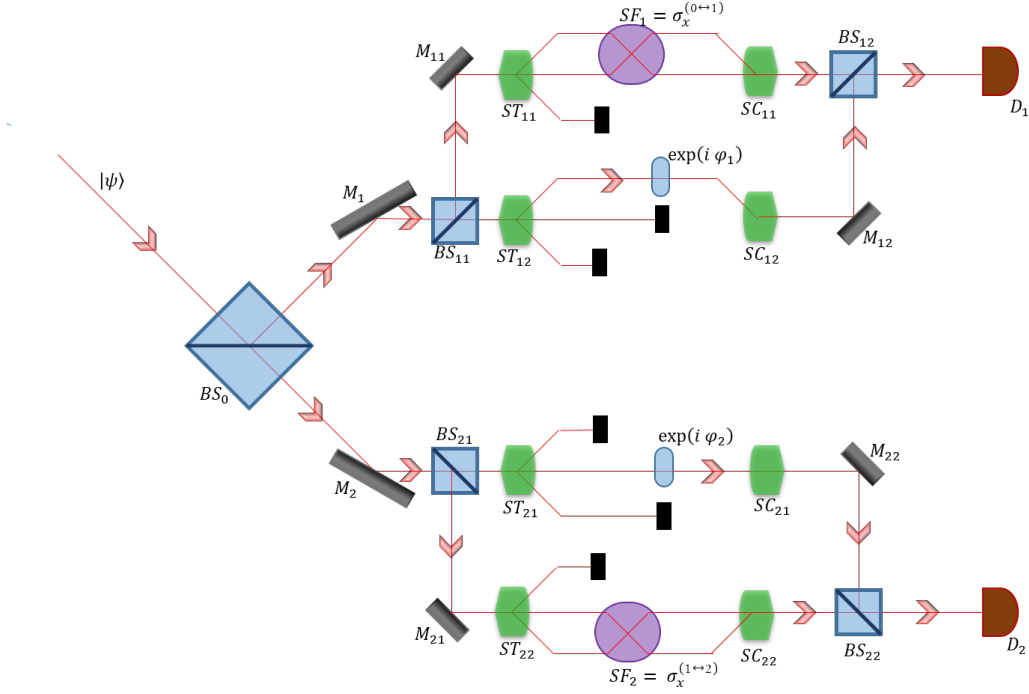

FIG. 9. Schematic of the generalization of the protocol to measure the quantum state of any qutrit. This method is scalable to any higher dimensional system.

On one arm of the MZI, we just add a phase shift $\exp(i\varphi_1)$ in spin mode $|0\rangle$ and block mode $|1\rangle$. The effective operator in this arm is then given by $\exp(i\varphi_1)\Pi_0^{(1)}$. In the other arm of the MZI, we use the spin flipper $(SF_1)$ to swap the mode $|0\rangle$ with $|1\rangle$. Finally, the spin modes in each arm are recombined using the spin combiner(reverse of tritter) $SC_{11}$ and $SC_{12}$ respectively. The two spatial modes are recombined using $BS_{12}$ and one of the ports of this MZI needs to be detected. Intensity $I_{d_1}$ is measured as a function of $\varphi_1$ to obtain visibility, phase shift and phase averaged intensity in the detector $D_1$. Similarly, the second MZI consist of $BS_{21}$ and $BS_{22}$ and acts on the spin subspace $\{|1\rangle, |2\rangle\}$. After the tritters $ST_{21}$ and $ST_{22}$, the spin mode $|0\rangle$ is blocked. On one arm we have the operator $\exp(i\varphi_2)\Pi_0^{(2)}$ and on the other arm we have the spin-flip operator that swaps mode $|1\rangle$ with $|2\rangle$. This scheme can be generalized with $d$-dimensions simply by blocking all other components after the spin splitter (ST) except the desired pair.

## XII. INFERRING BLOCH PARAMETERS

For the subspace spanned by $\{|k\rangle, |k+1\rangle\}$ with ($k \geq 1, \theta_d = 0$), we can write the modes as

$$|\psi\rangle_k^{(2;d)} = \left(\prod_{j=1}^{k-1} \sin\left(\frac{\theta_j}{2}\right) \exp(i\phi_j)\right) \begin{pmatrix} \cos\left(\frac{\theta_k}{2}\right) \\ \sin\left(\frac{\theta_k}{2}\right) \exp(i\phi_k) \cos\left(\frac{\theta_{k+1}}{2}\right) \end{pmatrix} \quad (29)$$

The first term is nothing but global phase multiplied with the amplitude of the vector in this subspace and hence the intensity modulation $I_k$ would not be affected by it but just scaled by the factor $\xi(k) = \prod_{j=1}^{k-1} \sin^2(\frac{\theta_j}{2})$.

Therefore, the intensity for the $k$-th subspace is given by

$$I_k = \frac{1}{16} \left( \prod_{j=1}^{k-1} \sin^2\left(\frac{\theta_j}{2}\right) \right) \left( 5 + \cos(\theta_k)(3 - \cos(\theta_{k+1})) + \cos(\theta_{k+1}) + 4\sin(\theta_k)\cos\left(\frac{\theta_{k+1}}{2}\right)\cos(\phi_k - \varphi_k) \right). \quad (30)$$

Hence, we would obtain the phase shift as $\Phi_k = \phi_k$.
The phase averaged intensity is obtained as

$$\bar{I}_k = \frac{1}{16} \left( \prod_{j=1}^{k-1} \sin^2\left(\frac{\theta_j}{2}\right) \right) \left( 5 + \cos(\theta_{k+1}) + \cos(\theta_k)(3 - \cos(\theta_{k+1})) \right) \quad (31)$$

Using the above equation, we can determine $\theta_k$ from $\bar{I}_k$ for a given $\theta_{k+1}$ and $\left( \prod_{j=1}^{k-1} \sin^2(\frac{\theta_j}{2}) \right)$.

From $(d-1)$ MZ interferometers we have $(d-1)$ measured values of $\bar{I}_k$ and need to infer $(d-1)$ values of $\theta_k$'s. The determination of the $\theta_k$'s is not straight forward because of the presence of $\xi(k) = \prod_{j=1}^{k-1} \sin^2(\frac{\theta_j}{2})$ term.

To eliminate the product of sines, we take the ratio of $\bar{I}_k$ to $\bar{I}_{k-1}$.

$$\frac{\bar{I}_k}{\bar{I}_{k-1}} = \frac{\sin^2(\frac{\theta_{k-1}}{2})(5 + \cos(\theta_{k+1}) + \cos(\theta_k)(3 - \cos(\theta_{k+1})))}{(5 + \cos(\theta_k) + \cos(\theta_{k-1})(3 - \cos(\theta_k)))} \quad (32)$$

However, now we have three variables $\theta_{k-1}$, $\theta_k$ and $\theta_{k+1}$ as opposed to one ratio. Since we know $\theta_d = 0$ because $\cos(\theta_d) = 1$, we can eliminate one variable in the ratio, but still have two variables left.

$$\frac{\bar{I}_{d-1}}{\bar{I}_{d-2}} = \frac{2\sin^2(\frac{\theta_{d-2}}{2})(3 + \cos(\theta_{d-1}))}{(5 + \cos(\theta_{d-1}) + \cos(\theta_{d-2})(3 - \cos(\theta_{d-1})))} \quad (33)$$

Now, we can make use of the visibility of the interference pattern for the $k$-th subspace as another known quantity to infer $\theta_{d-1}$ and $\theta_{d-2}$.

$$V_k = \frac{4\cos\left(\frac{\theta_{k+1}}{2}\right)\sin(\theta_k)}{5 + \cos(\theta_k)(3 - \cos(\theta_{k+1})) + \cos(\theta_{k+1})} \quad (34)$$

Thus, we have

$$V_{d-1} = \frac{2\sin(\theta_{d-1})}{3 + \cos(\theta_{d-1})} \quad (35)$$

We can obtain, $\theta_{d-1}$ from Eqn. (35) and then plug it in Eqn. (33) or in the expression $V_{d-2}$ to obtain $\theta_{d-2}$. Knowing $\theta_{d-2}$, we can infer $\theta_{d-3}$ from $V_{d-3}$ or the ratio $\frac{\bar{I}_{d-2}}{\bar{I}_{d-3}}$ and so on. In this sequential method we can obtain all the polar angles $\theta_k$'s.

*a. Alternative Method* Alternatively, we can solve for $\theta_1$ and $\theta_2$ using visibility $V_1$ and phase averaged intensity $\bar{I}_1$.

$$V_1 = \frac{4\cos\left(\frac{\theta_2}{2}\right)\sin(\theta_1)}{5 + \cos(\theta_1)(3 - \cos(\theta_2)) + \cos(\theta_2)} \quad (36)$$

$$\bar{I}_1 = \frac{1}{16}(5 + \cos(\theta_2) + \cos(\theta_1)(3 - \cos(\theta_2))) \quad (37)$$

We have to solve the above two equations simultaneously. The advantage is that once we know $\theta_1, \theta_2$, we can infer $\theta_3$ from $\bar{I}_2$ (or $V_2$). We only have to solve the simultaneous equations once and then use the recursive property of $\bar{I}_k$ and $V_k$ to infer other $\theta_k$'s. Note that, from above we may have multiple solutions in which case we have to resort to solving for $\theta_k$ and $\theta_{k+1}$ using $V_k$ and $\bar{I}_k$.

*b. Normalization* The projection of the state in the $k$-th subspace $|\psi\rangle_k^{(2;d)}$ is not normalized. Apriori, there is no need for normalization because the projection of a vector need not have pre-defined norm. Nevertheless, below we express the expectation values in normalized form.

$$\frac{\langle\psi|\sigma_-^{(k)}|\psi\rangle_k^{(2;d)}}{\langle\psi|\psi\rangle_k^{(2;d)}} = \frac{2\sin(\theta_k)e^{-i\phi_k}\cos\left(\frac{\theta_{k+1}}{2}\right)}{3 + \cos(\theta_{k+1}) + \cos(\theta_k)(1 - \cos(\theta_{k+1}))} \quad (38)$$

Thus, even in normalized form, the argument of the complex expectation value of the ladder operator $\sigma_-^{(k)}$ still gives us the relative phase in the $k$-th subspace. The above form assumes that the factors $\prod_{j=1}^{k-1} \sin^2\left(\frac{\theta_j}{2}\right)$ cancel out between numerator and denominator requiring that $\theta_j \neq 0 \ \forall j < k$. However, it should be pointed out that in the limiting case, the terms always cancel out. However, when $\theta_k = 0$ and $\theta_{k+1} = 0$ simultaneously, $|\psi\rangle_k^{(2;d)}$ cannot be normalized.

Similarly, we can express the average intensity in the normalized form as follows:

$$\bar{I}_k = \frac{5 + \cos(\theta_{k+1}) + \cos(\theta_k)(3 - \cos(\theta_{k+1}))}{4(3 + \cos(\theta_{k+1}) + \cos(\theta_k)(1 - \cos(\theta_{k+1})))} \quad (39)$$

The advantage of this normalized form is $\bar{I}_k$ now only contains the terms involving $\theta_k$ and $\theta_{k+1}$ and does not carry the effect of all $\theta_j$'s in the form of $\prod_{j=1}^{k-1} \sin^2\left(\frac{\theta_j}{2}\right)$. Therefore, we can directly compute $\theta_{d-1}$ using $\bar{I}_{d-1}$.

$$\bar{I}_{d-1} = \frac{5 + \cos(\theta_d) + \cos(\theta_{d-1})(3 - \cos(\theta_d))}{4(3 + \cos(\theta_{d-1}) + \cos(\theta_d) - \cos(\theta_{d-1})\cos(\theta_d))}$$

$$\quad (40)$$

$$= \frac{3 + \cos(\theta_{d-1})}{8} \quad (41)$$

Knowing $\theta_{d-1}$, we can know $\theta_{d-2}$ from $\bar{I}_{d-2}$ and then iteratively all the $\theta_k$'s can be found.

Additionally, we can make use of the visibility of the interference pattern for the $k$-th subspace as another known quantity to infer $\theta_{d-1}$ and $\theta_{d-2}$.

$$V_k = \frac{4\cos\left(\frac{\theta_{k+1}}{2}\right)\sin(\theta_k)}{5 + \cos(\theta_k)(3 - \cos(\theta_{k+1})) + \cos(\theta_{k+1})} \quad (42)$$

Thus, we have

$$V_{d-1} = \frac{2\sin(\theta_{d-1})}{3 + \cos(\theta_{d-1})} \ . \quad (43)$$

We can obtain, $\theta_{d-1}$ from Eqn. 43 and then plug it in the expression $V_{d-2}$ to obtain $\theta_{d-2}$. Knowing $\theta_{d-2}$, we can infer $\theta_{d-3}$ from $V_{d-3}$ and so on. In this sequential method we can obtain all the polar angles $\theta_k$'s.

## XIII. NO. OF INTERFEROMETERS NEEDED TO PERFORM QSI FOR QUDITS

We have established QSI with Mach Zehnder interferometry to show that the quantum state can be inferred from the interference pattern. The use of MZI in experiments is not trivial because of the need to stabilize them. Therefore, in the experiment, we have used displaced Sagnac interferometer, where there is no need for stabilization. Also, the use of interferometer in a non-collinear configuration gives us the intensity as function of phase difference directly without changing any settings in the experimental set-up. For $d$-dimensional pure qudit, it seems that we may need $d-1$ MZI interferometers and hence it would be experimentally challenging to set each one of them up with correct phase stabilization. Depending on the system, it could be experimentally easier to switch to other types of interferometers like Sagnac or double slit interferometer which do not require stabilization and hence are relatively easy to set up in practice.

However, it is not always necessary to have $d-1$ interferometers. To reconstruct a qudit state using QSI one needs more and more *interference patterns* as the dimension goes higher and higher. But the scaling is linear i.e., only $d-1$ interference patterns are required to infer the state of $d$-dimensional qudit. These interference patterns can be obtained by using only two interferometers if we observe the interference pattern on a camera (2D imaging sensor). The key idea is to stack the interference patterns vertically (assuming each interference pattern is formed along the horizontal) so that the same interferometer can be used for obtaining the required interference in each subspace.

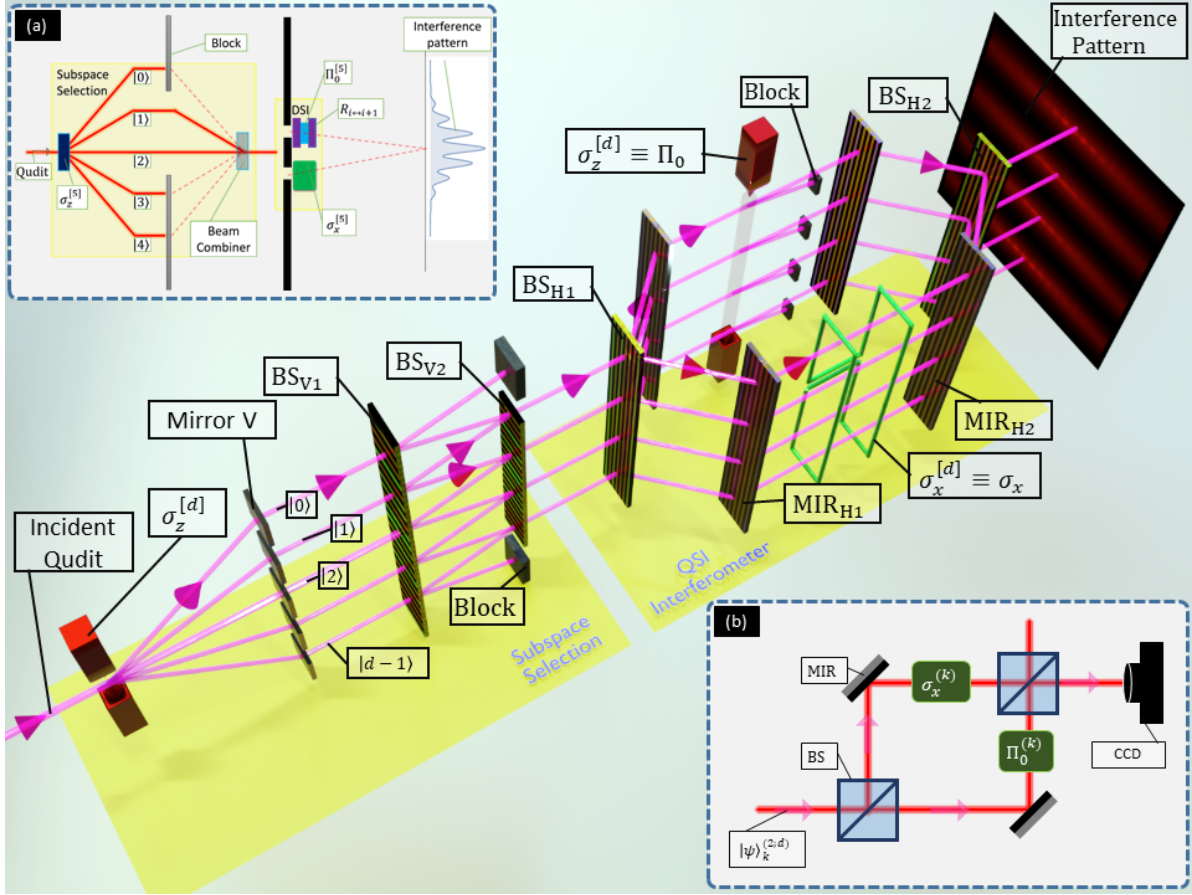

FIG. 10. The pure state of $d$-dimensional qudit can be inferred from $d-1$ interferograms obtained with the use of only two interferometers. The first interferometer is used to select the 2-dimensional subspaces on which QSI can be performed using the second interferometer. First the incident qudit is decomposed into eigenstates of $\sigma_z^{[d]}$. This could be implemented for spin degree of freedom by an inhomogeneous magnetic field along $\hat{z}$ so that the components split vertically. The Bragg mirrors make all the components parallel. The Bragg beam splitter $BS_{V_1}$ then splits each of the beam vertically with equal amplitudes. The beams from adjacent eigenstates $|k-1\rangle$ and $|k\rangle \ \forall \ k = 1, 2\ldots, d$ are combined using the Bragg beam combiner $BS_{V_2}$ after which we have the $d-1$ 2-dimensional subspaces. Here, the combination needs to be coherent and hence the spin splitter $\sigma_z^{[d]}$ and the beam combiner $BS_{V_2}$ forms the first interferometer with which $d-1$ subspaces can be created simultaneously by stacking the beams vertically. Half of the intensity of $|0\rangle$ and $|d-1\rangle$ is blocked so that each 2-dimensional subspace has *apriori* unbiased contribution from the two eigenstates. Next, we need to obtain the interferograms for each of the subspaces. As shown in the inset $(b)$, we need to construct a two path interferometer, such as the MZI, at which the state $|\psi\rangle_k^{2;d}$ in the $k$-th 2-dimensional subspace of the $d$-dimensional qudit is made incident. On one arm, we need the projector $\Pi_0^{(k)}$ for the $k$-th subspace and on the other arm we need to implement the evolution operator $\sigma_x^{(k)}$ for the $k$-th subspace. The Bragg beam splitter $BS_{H_1}$, which splits the beam horizontally with equal amplitudes, along with the beam combiner $BS_{H_2}$ forms the Mach Zehnder interferometer. The interferometer can be made non-collinear such that the interference pattern is obtained along the horizontal as shown in the figure. The $d-1$ interferograms can be obtained using the same interferometer as the subspaces are stacked vertically. The use same pair of beam splitter and combiner for all the $d-1$ subspaces implies that instead of stabilizing the otherwise $d-1$ MZIs now we need to stabilize only one. Also, the operator $\Pi_0^{(k)}$ for the $k$-th 2-dimensional subspace can be realized by simply using the $\sigma_z^{[d]}$ measurement operator for the qudit in one arm. This operator will split the subspace into two eigenstates. For the $k$-th subspace, the state $|k\rangle$ can be blocked and the state $|k-1\rangle$ can be allowed through the interferometer to effectively realize the operator $\Pi_0^{(k)}$. The evolution operator $\sigma_x^{(k)}$ for the $k$-th subspace is the $\sigma_x^{[d]}$ evolution operator for the $d$-dimensional qudit. Thus, the operations $R$ and $U$ to obtain the interferograms in all the 2-dimensional subspaces can have same physical implementation. Hence, without any change in settings, with only two interferometers - one used to prepare the 2-dimensional subspaces and the other used to perform QSI on the subspaces, we can reconstruct the pure state of a $d$-dimensional qudit. Although the above is discussed in the context of spin degree of freedom, the above can be achieved for most systems with suitable operators. If capturing a 2D image is not possible in a specific system, we can alternatively change the settings $d-1$ times in the two interferometers to obtain the $d-1$ interferograms as shown in $(a)$ - the details of which is described in the next Figure 11.

Although the scheme presented in Fig. 5 of the manuscript and Fig. 9 in this supplementary material is conceptual extension of QSI for qubits to higher dimensions it is lossy as in each interferometer we discard all particles that do not belong to the desired subspace. Since we choose only 2-dimensional subspace, only $\mathcal{O}(2/d)$ particles are used for QSI. However, if we make the modification as shown above in Fig. 10, not only the requirement of the interferometer goes down to 2 for all $d > 2$ but also QSI becomes more efficient. Before the second interferometer, only half of the particles belonging to the extreme eigenstates of $\sigma_z^{[d]}$ are blocked. Thus, $\mathcal{O}\left(\frac{d-1}{d}\right)$ particles are used. Thus, the losses are negligible for higher dimensional systems.

Note that when the input is restricted to the subspace $\{|k\rangle, |k+1\rangle\}$, the $\sigma_x^{[d]}$ operation for in the $d = 5$ dimensional Hilbert space also acts as the $\sigma_x^{(k)}$ operator for the $k$-th subspace.

The matrix element of $\sigma_x^{[d]}$ operator is given by

$$\sigma_x\Big|_{i,j}^{[d]} = (\delta_{i+1,j} + \delta_{i,j+1})\sqrt{(s+1)(i+j-1) - ij} \quad (44)$$

Here, the dimension of the Hilbert space is given by $d = 2s+1$ and $i,j = 1,2,\ldots d$. For the $k$-th subspace, we have $i = k, k+1$ and $j = k, k+1$. Thus, the corresponding matrix elements are given as follows

$$\sigma_x\Big|_{k,k}^{[d]} = (\delta_{k+1,k} + \delta_{k,k+1})\sqrt{(s+1)(2k-1) - k^2}$$
$$= 0 \quad (45)$$

$$\sigma_x\Big|_{k,k+1}^{[d]} = (\delta_{k+1,k+1} + \delta_{k,k+2})\sqrt{(s+1)(2k) - k(k+1)}$$
$$= \sqrt{k(d-k)} \quad (46)$$

$$\sigma_x\Big|_{k+1,k}^{[d]} = (\delta_{k+2,k} + \delta_{k+1,k+1})\sqrt{(s+1)(2k) - k(k+1)}$$
$$= \sqrt{k(d-k)} \quad (47)$$

$$\sigma_x\Big|_{k+1,k+1}^{[d]} = (\delta_{k+2,k+1} + \delta_{k+1,k+2})\sqrt{(s+1)(2k+1) - (k+1)^2}$$
$$= 0 \quad (48)$$

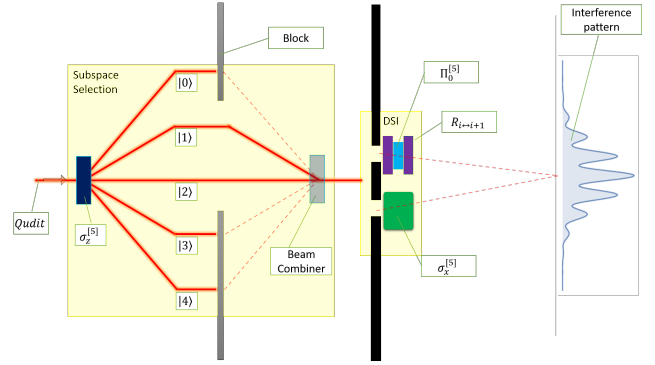

FIG. 11. Schematic of Quantum State Interferography for a generic $d(= 5)$-dimensional qudit with slit interferomters: The stream of particles is first resolved into the $d(= 5)$ eigenstates of the $\sigma_z^{[d=5]}$ operator that acts on the $d(= 5)$-dimensional Hilbert space. Any two of these $d(= 5)$ beams (say $|1\rangle$ and $|2\rangle$), can be allowed to be combined and rest can be blocked. This needs to be coherent combination and effectively the beam combiner along with $\sigma_z^{[d=5]}$ would be just one interferometer (say of the type Sagnac interferometer). To perform, QSI on the selected subspace (k), we now need only one more double slit interferometer. On one of the slits, we have the $\sigma_x^{(k)}$. This can simply be the $\sigma_x^{[5]}$ operator on the $d = 5$ dimensional Hilbert space as shown below. On the other slit, we need to have the projector $\Pi_0^{(k)}$, which can be constructed from the projector $\Pi_0^{[5]}$ by rotation of basis using operators $R_{i \leftrightarrow i+1}$. These rotation operations can be changed to realize different $\Pi_0^{(k)}$ for the choice of subspace $\{|k\rangle, |k+1\rangle\}$. Hence, overall for arbitrarily high $d$-dimensional qudits, only two interferometers are needed to be set up, one for combining the beams in a selected subspace (say Sagnac) and the other being double slit interferometer. The cost, however is that, we need to change the block to allow a certain combination $d-1$ times and for each such combination the projector needs to be rotated so that it is set to $\Pi_0^{(k)}$.

Thus, the $\sigma_x^{(k)}$ operator for the $k$-th subspace can be obtained by the action of $\sigma_x^{[d]}$ on the subspace.

$$\sigma_x^{(k)} = \begin{pmatrix} \sigma_x\big|_{k,k}^{[d]} & \sigma_x\big|_{k,k+1}^{[d]} \\ \sigma_x\big|_{k+1,k}^{[d]} & \sigma_x\big|_{k+1,k+1}^{[d]} \end{pmatrix} \quad (49)$$

$$= \sqrt{k(d-k)} \begin{pmatrix} 0 & 1 \\ 1 & 0 \end{pmatrix} \quad (50)$$

As we can see from the example below, the highlighted

2-dimensional matrices are a scalar times the $\sigma_x$ operation in the respective 2-dimensional subspace. This holds true for all sequential pairwise two-dimensional subspaces for arbitrary $d$-dimensions.

$$\sigma_x^{[5]} = \begin{pmatrix} 0 & 2 & 0 & 0 & 0 \\ 2 & 0 & \sqrt{6} & 0 & 0 \\ 0 & \sqrt{6} & 0 & \sqrt{6} & 0 \\ 0 & 0 & \sqrt{6} & 0 & 2 \\ 0 & 0 & 0 & 2 & 0 \end{pmatrix}$$

Thus, we have shown that, QSI can be performed with just two interferometers for any $d$ dimensions. The operators for the subspace can be realized with the operators available for $d$-dimensional Hilbert space.

## XIV. QSI FOR BIPARTITE SYSTEMS

In the manuscript and in the previous sections of this supplementary, we have discussed QSI to determine arbitrary (pure or mixed) states of qubits and pure states of qudits. In this section, we expand the scope of QSI to determine the state of bipartite systems. We shall work out the scheme for a pair of photons generated by Spontaneous Parametric Down-Conversion (SPDC) process but the method shall be applicable to all bipartite systems once we identify the realization of relevant operators. We shall not use any global operation, so that the parties, say Alice (with signal particle A) and Bob (with idler particle B), can perform local single qubit operations and later by classical communication i.e., post-processing with coincidence logic, determine the state of the bipartite system. For this, both Alice and Bob need to perform the single qubit quantum state interferography on their respective signal and idler photons and record the time-stamps of the particles forming the interference pattern. One of them, here Bob, needs to perform QSI with the heralded $B$ particles subject to $A$ particles being projected on to $|H\rangle$ which can be achieved by blocking the arm containing $\sigma_x$ in Alice's setup. Bob can extract the heralded intensity pattern by correlating time-stamps of signal photons projected to $|H\rangle$ with idler. Thus, in total, for the QSI of bipartite system, Alice and Bob would need just two-experimental settings to obtain 3 interference patterns - singles with particle A, singles with particle B and heralded interference pattern of particle B conditioned to particle A being projected onto $|H\rangle$. Just

for comparison, Quantum State Tomography for bipartite qubit state assuming that the state is pure would require 9 measurement settings. Also, as mentioned in the main text, performing QSI on either particles $A$ or $B$ to obtain the unheralded i.e. singles interference pattern can give us the reduced density matrix. This is useful to quantify entanglement with a single setup.

In general, the state of a bipartite system can be written as

$$|\Psi\rangle_{AB} = \alpha_1 |HH\rangle_{AB} + \alpha_2 |HV\rangle_{AB} + \alpha_3 |VH\rangle_{AB} + \alpha_4 |VV\rangle_{AB} \quad (51)$$

where, $|HH\rangle_{AB} = |H\rangle_A |H\rangle_B$ etc. We have 4 complex numbers as the coefficients, constrained to the normalization condition that $\sum_{i=1}^{4} |\alpha_i|^2 = 1$. Also, we can ignore the global phase. Thus, we have to determine 6 real quantities from the experiment.

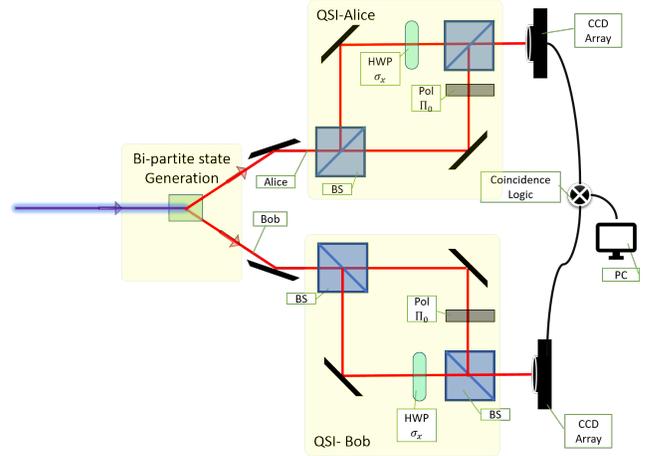

FIG. 12. QSI for bipartite system: Alice and Bob performs single qubit QSI on their respective particles.

We can collect $A$'s polarization so that it helps us to parameterize the state in terms of angles for a single qubit Bloch sphere.

$$\begin{aligned} |\Psi\rangle_{AB} &= \alpha_1 |HH\rangle_{AB} + \alpha_2 |HV\rangle_{AB} + \alpha_3 |VH\rangle_{AB} + \alpha_4 |VV\rangle_{AB} \\ &= |H\rangle_A (\alpha_1 |H\rangle_B + \alpha_2 |V\rangle_B) + |V\rangle_A (\alpha_3 |H\rangle_B + \alpha_4 |V\rangle_B) \end{aligned} \quad (52)$$

Once, the signal is projected to $|H\rangle_A$, the idler state becomes $|\psi\rangle_B^H = (\alpha_1 |H\rangle_B + \alpha_2 |V\rangle_B)$. Note that if we

start with a pure bipartite state $|\Psi\rangle_{AB}$, the state of the idler also reduces to a pure state once the signal is projected to a particular pure polarization state. The term $|\psi\rangle_B^H$ appears to have the same form as a single qubit, but we have to consider that it is not normalized to 1. Let the normalization be $\gamma$ which is the probability of the particle $A$ to be projected in state $|H\rangle_A$. If Alice just blocks the arm with $\sigma_x$ operation to effectively perform this projection $\gamma = 4P_{d_A}^H$. Here, $P_{d_A}^H$ is the probability of Alice detecting the horizontally projected particles in his CCD Array. The factor 4 is introduced to compensate for the particles not collected due to the two beam splitters. Then $\alpha_1$ and $\alpha_2$ can be written in terms of the Bloch sphere angles $\theta_H$ and $\phi_H$.

$$\alpha_1 = \sqrt{\gamma} \cos\left(\frac{\theta_H}{2}\right) \tag{53}$$

$$\alpha_2 = \sqrt{\gamma} \sin\left(\frac{\theta_H}{2}\right) e^{i\phi_H} \tag{54}$$

The phase of the complex number $\alpha_1$ is absorbed as the global phase. The subscript $H$ reminds us that the state in the Bloch sphere represents the state for the idler qubit when signal is projected to $|H\rangle_A$. Similarly, when the signal is projected to $|V\rangle_A$, we have the parameterization of the reduced idler state $|\psi\rangle_B^V$ as follows:

$$\alpha_3 = \sqrt{1-\gamma} e^{i\phi_r} \cos\left(\frac{\theta_V}{2}\right) \tag{55}$$

$$\alpha_4 = \sqrt{1-\gamma} e^{i\phi_r} \sin\left(\frac{\theta_V}{2}\right) e^{i\phi_V} \tag{56}$$

Here, $\phi_r$ is the relative phase between the Bloch vectors associated with $|\psi\rangle_B^H$ and $|\psi\rangle_B^V$, which are normalized to $\gamma$ and $1-\gamma$ respectively. We of course can write $\gamma = \cos^2(\theta_r)$ to explicitly make $\gamma \leq 1$ which helps in algebraic simplification later. Consequently, we substitute $\sqrt{1-\gamma} = \sin(\theta_r)$. Thus, any pure bipartite qubit state can be written in terms of the 6 real parameters $\theta_H, \phi_H, \theta_V, \phi_V, \theta_r$ and $\phi_r$. All the $\theta$'s belong to $[0, \pi)$ and all the $\phi$'s belong to $(-\pi, \pi]$.

$$|\Psi\rangle_{AB} = \begin{pmatrix} \cos(\theta_r)\cos\left(\frac{\theta_H}{2}\right) \\ \cos(\theta_r)\sin\left(\frac{\theta_H}{2}\right)e^{i\phi_H} \\ \sin(\theta_r)e^{i\phi_r}\cos\left(\frac{\theta_V}{2}\right) \\ \sin(\theta_r)e^{i\phi_r}\sin\left(\frac{\theta_V}{2}\right)e^{i\phi_V} \end{pmatrix} \tag{57}$$

When Alice projects the particle $A$ to $|H\rangle_A$, the state of $B$ becomes,

$$|\psi\rangle_B^H = \alpha_1 |H\rangle_B + \alpha_2 |V\rangle_B \tag{58}$$
$$= \begin{pmatrix} \cos(\theta_r)\cos\left(\frac{\theta_H}{2}\right) \\ \cos(\theta_r)\sin\left(\frac{\theta_H}{2}\right)e^{i\phi_H} \end{pmatrix} = \cos(\theta_r) \begin{pmatrix} \cos\left(\frac{\theta_H}{2}\right) \\ \sin\left(\frac{\theta_H}{2}\right)e^{i\phi_H} \end{pmatrix} \tag{59}$$

Here, $\cos(\theta_r)$ is the global factor and would only affect the normalization of the state $|\psi\rangle_B^H$. Since Alice can determine $\theta_r$ from $P_{d_A}^H$, Bob can use it to suitably normalize his state from the observed intensity.

The evolution of state $|\psi\rangle_B^H$ through the QSI setup of Bob to one of the output port of the second beam splitter with the CCD Array, can be described by using the effective evolution operator $\mathcal{E}$ given by

$$\mathcal{E} = \frac{1}{2}(R\exp(i\varphi) + U) = \frac{1}{2}\begin{pmatrix} e^{i\varphi} & 1 \\ 1 & 0 \end{pmatrix}. \tag{60}$$

Here, $\varphi$ is the phase difference between the two arms of the interferometer. This evolution operator is non-unitary because of losses due to the operator $R$ and considering the fact that we are detecting in only one port of the second beam splitter of the interferometer. Note that $\mathcal{E}$ is only employed here as a short-cut to the detailed derivation done in Sec II in this supplementary.

The interferometer can be made non-collinear so that we get intensity as a function of phase difference $\varphi$ in the CCD array. We assume that the CCD array can be gated or has time-stamps so that correlation can be obtained with the detected particles of Alice. The heralded intensity pattern obtained by Bob conditioned to Alice detecting particle $A$ in $|H\rangle_A$ is obtained as

$$I_B^H = |\mathcal{E} |\psi\rangle_B^H|^2 \tag{61}$$
$$= \frac{1}{8}\cos^2(\theta_r)\left(3 + \cos(\theta_H) + 2\sin(\theta_H)\cos(\varphi - \phi_H)\right) \tag{62}$$

The above equation is similar to the Eqn. 11 except the additional scaling factor $\cos^2(\theta_r)$ (apart from of course $\mu = 1$ due to assumption of pure state).

The phase shift of the interference pattern $I_B^H(\varphi)$ can be obtained by finding the $\varphi$ at which $I_B^H(\varphi)$ is maximum. This is obtained by solving for $\varphi$ in the equation $\frac{\partial}{\partial\varphi}I_B^H(\varphi) = 0$ and ensuring that $\frac{\partial^2}{\partial\varphi^2}I_B^H(\varphi) < 0$. We obtain the phase shift of the interference pattern $\Phi_B^H$ as

$\phi_H$. Thus the real quantity $\phi_H$ is directly obtained from the phase shift of the interference pattern $I_B^H$.

The phase averaged intensity is obtained as

$$\overline{I_B^H} = \frac{1}{8}\cos^2(\theta_r)(3 + \cos(\theta_H)) \tag{63}$$

Since Alice can determine $\cos^2(\theta_r)$ from the probability of particle $A$ detected in state $|H\rangle_A$, Bob can use that information to determine $\theta_H$ from the above expression.

Thus, from the heralded interference pattern, Alice and Bob can determine 3 parameters $\theta_r, \theta_H$ and $\phi_H$. Now, if Bob considers all the particles detected in the CCD without heralding, the reduced state of particle $B$ is, in general, mixed. Thus, we need to consider the density matrix $\rho_{AB} = |\Psi\rangle_{AB}\langle\Psi|_{AB}$ from which we can obtain the reduced density matrix for particle $B$ after performing partial trace over $A$, $\rho_B = \text{Tr}_A(\rho_{AB})$.

$$\rho_{AB} = \frac{1}{2}\begin{pmatrix} 2c_{\frac{\theta_H}{2}}^2 c_{\theta_r}^2 & e^{-i\phi_H}s_{\theta_H}c_{\theta_r}^2 & e^{-i\phi_r}c_{\frac{\theta_H}{2}}s_{2\theta_r}c_{\frac{\theta_V}{2}} & c_{\frac{\theta_H}{2}}s_{2\theta_r}s_{\frac{\theta_V}{2}}e^{-i(\phi_r+\phi_V)} \\ e^{i\phi_H}s_{\theta_H}c_{\theta_r}^2 & 2s_{\frac{\theta_H}{2}}^2 c_{\theta_r}^2 & s_{\frac{\theta_H}{2}}s_{2\theta_r}c_{\frac{\theta_V}{2}}e^{i(\phi_H-\phi_r)} & s_{\frac{\theta_H}{2}}s_{2\theta_r}s_{\frac{\theta_V}{2}}e^{i(\phi_H-\phi_r-\phi_V)} \\ e^{i\phi_r}c_{\frac{\theta_H}{2}}s_{2\theta_r}c_{\frac{\theta_V}{2}} & s_{\frac{\theta_H}{2}}s_{2\theta_r}c_{\frac{\theta_V}{2}}e^{i(\phi_r-\phi_H)} & 2s_{\theta_r}^2 c_{\frac{\theta_V}{2}}^2 & e^{-i\phi_V}s_{\theta_r}^2 s_{\theta_V} \\ c_{\frac{\theta_H}{2}}s_{2\theta_r}s_{\frac{\theta_V}{2}}e^{i(\phi_r+\phi_V)} & s_{\frac{\theta_H}{2}}s_{2\theta_r}s_{\frac{\theta_V}{2}}e^{i(-\phi_H+\phi_r+\phi_V)} & e^{i\phi_V}s_{\theta_r}^2 s_{\theta_V} & 2s_{\theta_r}^2 s_{\frac{\theta_V}{2}}^2 \end{pmatrix} \tag{64}$$

Here, $c_\theta$ and $s_\theta$ are $\cos(\theta)$ and $\sin(\theta)$ functions respectively.

The reduced density matrix $\rho_B$ is obtained as

$$\rho_B = \text{Tr}_A(\rho_{AB}) \tag{65}$$

$$= \frac{1}{2}\begin{pmatrix} 2\left(\cos^2\left(\frac{\theta_H}{2}\right)\cos^2(\theta_r) + \sin^2(\theta_r)\cos^2\left(\frac{\theta_V}{2}\right)\right) & e^{-i\phi_H}\sin(\theta_H)\cos^2(\theta_r) + e^{-i\phi_V}\sin^2(\theta_r)\sin(\theta_V) \\ e^{i\phi_H}\sin(\theta_H)\cos^2(\theta_r) + e^{i\phi_V}\sin^2(\theta_r)\sin(\theta_V) & 2\left(\sin^2\left(\frac{\theta_H}{2}\right)\cos^2(\theta_r) + \sin^2(\theta_r)\sin^2\left(\frac{\theta_V}{2}\right)\right) \end{pmatrix} \tag{66}$$

The unconditional interference pattern is obtained from $\rho_B$ as follows:

$$I_B = \text{Tr}(\mathcal{E}\rho_B\mathcal{E}^\dagger) \tag{67}$$

$$= \frac{1}{8}\left(3 + \cos^2(\theta_r)(2\sin(\theta_H)\cos(\varphi - \phi_H) + \cos(\theta_H)) + \sin^2(\theta_r)(2\sin(\theta_V)\cos(\varphi - \phi_V) + \cos(\theta_V))\right) \tag{68}$$

The phase averaged intensity is computed to be

$$\overline{I_B} = \frac{1}{8}(3 + \cos(\theta_H)\cos^2(\theta_r) + \cos(\theta_V)\sin^2(\theta_r)) \tag{69}$$

Since, $\theta_r$ and $\theta_H$ are already known, we can infer $\theta_V$ from the above equation. Once the phase shift of the interference pattern is experimentally determined to be $\Phi_B$, the quantity $\phi_V$ can be inferred from the relation below:

$$\frac{\sin(\Phi_B - \phi_V)}{\sin(\Phi_B - \phi_H)} = -\cot^2(\theta_r)\frac{\sin(\theta_H)}{\sin(\theta_V)} \tag{70}$$

$$\Rightarrow \phi_V = \Phi_B + \arcsin\left(\cot^2(\theta_r)\frac{\sin(\theta_H)}{\sin(\theta_V)}\sin(\Phi_B - \phi_H)\right) \tag{71}$$

The above relation is obtained from $\left.\frac{\partial}{\partial\varphi}I_B(\varphi)\right|_{\varphi=\Phi_B} = 0$. Since, $\theta_r$, $\theta_H$, $\theta_V$ and $\phi_H$ are obtained earlier, we can determine $\phi_V$ from the above expression. Care should be taken to verify that $\phi_V$ maximizes $I_B$ for the values

of $\theta_H$ and $\phi_H$ obtained earlier. If $I_B$ gets minimized instead, $\pi$ must be added for consistency.

So far, we have determined all the quantities except $\phi_r$. Although, without $\phi_r$, the state $|\Psi\rangle_{AB}$ is not completely determined, we can infer a lot of properties such as whether the pure state $|\Psi\rangle_{AB}$ is entangled or not. The von Neumann entropy $S$ of the reduced state $\rho_B$ is a unique measure of entanglement for bipartite pure states. Since, $\phi_r$ does not appear in $\rho_B$, the entanglement quantified by the measure $E$.

$$E(\rho_{AB}) = S(\rho_B) = -\operatorname{Tr}(\rho_B \log(\rho_B)) \quad (72)$$

Next, for complete state reconstruction, we need to perform QSI on the unheralded $A$ particles in Alice's setup. The reduced density matrix for Alice is obtained as

$$\rho_A = \operatorname{Tr}_B(\rho_{AB}) \quad (73)$$

$$= \frac{1}{2}\begin{pmatrix} 2c_{\theta_r}^2 & e^{-i\phi_r}s_{2\theta_r}\left(c_{\frac{\theta_H}{2}}c_{\frac{\theta_V}{2}} + s_{\frac{\theta_H}{2}}s_{\frac{\theta_V}{2}}e^{i(\phi_H-\phi_V)}\right) \\ e^{i\phi_r}s_{2\theta_r}\left(c_{\frac{\theta_H}{2}}c_{\frac{\theta_V}{2}} + s_{\frac{\theta_H}{2}}s_{\frac{\theta_V}{2}}e^{-i(\phi_H-\phi_V)}\right) & 2s_{\theta_r}^2 \end{pmatrix} \quad (74)$$

The intensity as a function of $\varphi$ is given by

$$I_A = \operatorname{Tr}(\mathcal{E}\rho_A\mathcal{E}^\dagger) \quad (75)$$
$$= \frac{1}{8}\left(3 + \cos(2\theta_r) + 2\sin(2\theta_r)\left(\sin\left(\frac{\theta_H}{2}\right)\sin\left(\frac{\theta_V}{2}\right)\cos(\varphi+\phi_H-\phi_r-\phi_V) + \cos\left(\frac{\theta_H}{2}\right)\cos\left(\frac{\theta_V}{2}\right)\cos(\varphi-\phi_r)\right)\right) \quad (76)$$

Once we obtain the phase shift of the interference pattern by experimentally finding $\varphi = \Phi_A$ for which $I_A$ is maximum, we can determine $\phi_r$ by considering the relation $\left.\frac{\partial}{\partial \varphi} I_A(\varphi)\right|_{\varphi=\Phi_A} = 0$ since we know all other parameters.

$$\frac{\sin(\Phi_A - \phi_r)}{\sin(\Phi_A + \phi_H - \phi_V - \phi_r)} = -\tan\left(\frac{\theta_H}{2}\right)\tan\left(\frac{\theta_V}{2}\right) \quad (77)$$

$$\Rightarrow \phi_r = \Phi_A + \cot^{-1}\left(\cot(\phi_H - \phi_V) + \frac{\cot\left(\frac{\theta_H}{2}\right)\cot\left(\frac{\theta_V}{2}\right)}{\sin(\phi_H - \phi_V)}\right) \quad (78)$$

The above expression must be picked with appropriate signs which maximizes $I_A$ for a given set values of other parameters obtained earlier.

In summary, without changing the experimental settings, we can determine the quantities $\theta_r, \theta_H, \theta_V, \phi_H$ and $\phi_V$ from the two intensity patterns obtained by Bob in the same setup - one with $B$ particles heralded to Alice's $|H\rangle_A$ projection and the other with the unheralded $B$ particles. Only, $\phi_r$ cannot be determined from the above procedure and thus it would not serve us complete reconstruction of the pure state.

Nevertheless, we can quantify the entanglement of the bipartite pure state in a single setup. With alternative procedures like quantum state tomography one would require 9 measurements. For Bell inequality violation, one would need at least 3 measurement settings if the basis is known else the optimization procedure require many more measurements. For complete reconstruction, Alice need to perform another measurement i.e., record the interference pattern by performing QSI on qubit A. Hence, with 3 interference patterns from two experimental settings a pure bipartite qubit state can be reconstructed.

[1] Y. Aharonov, D. Z. Albert, and L. Vaidman, Phys. Rev. Lett. **60**, 1351 (1988).
[2] A. K. Pati, U. Singh, and U. Sinha, Phys. Rev. A **92**, 052120 (2015).
[3] G. Nirala, S. N. Sahoo, A. K. Pati, and U. Sinha, Phys. Rev. A **99**, 022111 (2019).
[4] Such custom designed slits have been in use in experiments such as S. P. Walborn, M. O. Terra Cunha, S. Padua, and C. H. Monken *Phys. Rev. A 65, 033818* (2002).
[5] K. Banaszek, M. Cramer, and D. Gross, New Journal of Physics **15**, 125020 (2013).
[6] D. F. V. James, P. G. Kwiat, W. J. Munro, and A. G. White, Physical Review A **64**, 052312 (2001).
[7] L. E. Blumenson, The American Mathematical Monthly **67**, 63 (1960).


\end{document}